\documentclass[aps,prd,twocolumn,showpacs,superscriptaddress,nofootinbib,floatfix,10pt]{revtex4-2}  %

\usepackage{graphicx}  
\usepackage{adjustbox} 
\usepackage{dcolumn}   
\usepackage{bm}        
\usepackage{amssymb}   
\usepackage{xcolor}
\usepackage{amsmath}
\usepackage{bm}
\usepackage{array}

\newcommand{\minerva}{MINERvA}

\usepackage[normalem]{ulem}
\usepackage{placeins}

\usepackage{appendix}

\newif\ifdraftmode \draftmodefalse     

\usepackage{xcolor}
\usepackage{soul}
\usepackage{tabularx}   
\usepackage{booktabs}   
\newcommand{\numuC}{$\nu_\mu$-carbon }
\newcommand{\genieTwo}{GENIE v2.12.6}
\newcommand{\genieThree}{GENIE v3.04.00 AR23\_20i\_00\_000}
\newcommand{\commStrike}[1]{}
\newcommand{\commComment}[1]{}
\newcommand{\AJBgeneseoPaperOnly}[1]{}
\begin{document}
\title{Boosted decision tree reweighting of simulated neutrino interactions for ${\cal O}(1)$~GeV neutrino cross section measurements}

\newcommand{\Rutgers}{Rutgers, The State University of New Jersey, Piscataway, New Jersey 08854, USA}
\newcommand{\Hampton}{Hampton University, Dept. of Physics, Hampton, VA 23668, USA}
\newcommand{\Dortmund}{Institute of Physics, Dortmund University, 44221, Germany }
\newcommand{\Otterbein}{Department of Physics, Otterbein University, 1 South Grove Street, Westerville, OH, 43081 USA}
\newcommand{\JMU}{James Madison University, Harrisonburg, Virginia 22807, USA}
\newcommand{\Florida}{University of Florida, Department of Physics, Gainesville, FL 32611}
\newcommand{\UCIrvine}{Department of Physics and Astronomy, University of California, Irvine, Irvine, California 92697-4575, USA}
\newcommand{\CBPF}{Centro Brasileiro de Pesquisas F\'{i}sicas, Rua Dr. Xavier Sigaud 150, Urca, Rio de Janeiro, Rio de Janeiro, 22290-180, Brazil}
\newcommand{\PUCP}{Secci\'{o}n F\'{i}sica, Departamento de Ciencias, Pontificia Universidad Cat\'{o}lica del Per\'{u}, Apartado 1761, Lima, Per\'{u}}
\newcommand{\INRM}{Institute for Nuclear Research of the Russian Academy of Sciences, 117312 Moscow, Russia}
\newcommand{\Jlab}{Jefferson Lab, 12000 Jefferson Avenue, Newport News, VA 23606, USA}
\newcommand{\Pittsburgh}{Department of Physics and Astronomy, University of Pittsburgh, Pittsburgh, Pennsylvania 15260, USA}
\newcommand{\Guanajuato}{Campus Le\'{o}n y Campus Guanajuato, Universidad de Guanajuato, Lascurain de Retana No. 5, Colonia Centro, Guanajuato 36000, Guanajuato M\'{e}xico.}
\newcommand{\Athens}{Department of Physics, University of Athens, GR-15771 Athens, Greece}
\newcommand{\Tufts}{Physics Department, Tufts University, Medford, Massachusetts 02155, USA}
\newcommand{\WM}{Department of Physics, William \& Mary, Williamsburg, Virginia 23187, USA}
\newcommand{\FNAL}{Fermi National Accelerator Laboratory, Batavia, Illinois 60510, USA}
\newcommand{\Purdue}{Department of Chemistry and Physics, Purdue University Calumet, Hammond, Indiana 46323, USA}
\newcommand{\MCLA}{Massachusetts College of Liberal Arts, 375 Church Street, North Adams, MA 01247}
\newcommand{\UMD}{Department of Physics, University of Minnesota -- Duluth, Duluth, Minnesota 55812, USA}
\newcommand{\Northwestern}{Northwestern University, Evanston, Illinois 60208}
\newcommand{\UNI}{Facultad de Ciencias F\'{i}sicas, Universidad Nacional Mayor de San Marcos, CP 15081, Lima, Per\'{u}}
\newcommand{\Rochester}{Department of Physics and Astronomy, University of Rochester, Rochester, New York 14627 USA}
\newcommand{\Austin}{Department of Physics, University of Texas, 1 University Station, Austin, Texas 78712, USA}
\newcommand{\USM}{Departamento de F\'{i}sica, Universidad T\'{e}cnica Federico Santa Mar\'{i}a, Avenida Espa\~{n}a 1680 Casilla 110-V, Valpara\'{i}so, Chile}
\newcommand{\Geneva}{University of Geneva, 1211 Geneva 4, Switzerland}
\newcommand{\Chicago}{Enrico Fermi Institute, University of Chicago, Chicago, IL 60637 USA}
\newcommand{\hired}{}
\newcommand{\OregonState}{Department of Physics, Oregon State University, Corvallis, Oregon 97331, USA}
\newcommand{\oxford}{Oxford University, Department of Physics, Oxford, OX1 3PJ United Kingdom}
\newcommand{\umiss}{University of Mississippi, Oxford, Mississippi 38677, USA}
\newcommand{\upenn}{Department of Physics and Astronomy, University of Pennsylvania, Philadelphia, PA 19104}
\newcommand{\AMU}{Department of Physics, Aligarh Muslim University, Aligarh, Uttar Pradesh 202002, India}
\newcommand{\wroclaw}{University of Wroclaw, plac Uniwersytecki 1, 50-137 Wroa\l{}aw, Poland}
\newcommand{\Mohali}{Department of Physical Sciences, IISER Mohali, Knowledge City, SAS Nagar, Mohali - 140306, Punjab, India}
\newcommand{\CINVESTAV}{Departamento de Fisica Col. San Pedro Zacatenco, 07360 Mexico, DF, Av. Instituto PolitÃ©cnico Nacional, Mexico}
\newcommand{\york}{York University, Department of Physics and Astronomy, Toronto, Ontario, M3J 1P3 Canada}
\newcommand{\ND}{Department of Physics and Astronomy, University of Notre Dame, Notre Dame, Indiana 46556, USA}
\newcommand{\ICL}{The Blackett Laboratory,  Imperial College London,  London SW7 2BW, United Kingdom}
\newcommand{\warwick}{Department of Physics, University of Warwick, Coventry, CV4 7AL, UK}
\newcommand{\qmul}{G O Jones Building, Queen Mary University of London, 327 Mile End Road, London E1 4NS, UK}

\newcommand{\ricfregianThanks}{now at Department of Physics and Astronomy, University of California at Davis, Davis, CA 95616, USA}
\newcommand{\kleykampThanks}{now at Department of Physics and Astronomy, University of Mississippi, Oxford, MS 38677}
\newcommand{\adrianThanks}{Now at Department of Physics, Drexel University, Philadelphia, Pennsylvania 19104, USA}
\newcommand{\byaeggyThanks}{Now at Department of Physics, University of Cincinnati,  Cincinnati, Ohio 45221, USA}
\newcommand{\lazazuetareyesThanks}{now at Syracuse University, Syracuse, NY 13244, USA}


\author{Z.~Lin}                           \affiliation{\Rochester}
\author{S.~Akhter}                        \affiliation{\AMU}
\author{Z.~~Ahmad~Dar}                    \affiliation{\WM}  \affiliation{\AMU}
\author{N.S.~Alex}                        \affiliation{\Rochester}
\author{M.~Betancourt}                    \affiliation{\FNAL}
\author{S.~Boyd}                          \affiliation{\warwick}  \affiliation{\Pittsburgh}
\author{H.~Budd}                          \affiliation{\Rochester}
\author{G.~Caceres}\thanks{\ricfregianThanks}  \affiliation{\CBPF}
\author{G.A.~D\'{i}az~}                   \affiliation{\FNAL}  \affiliation{\Rochester}
\author{J.~Felix}                         \affiliation{\Guanajuato}
\author{L.~Fields}                        \affiliation{\ND}
\author{A.M.~Gago}                        \affiliation{\PUCP}
\author{P.K.Gaur}                         \affiliation{\AMU}
\author{S.M.~Gilligan}                    \affiliation{\OregonState}
\author{R.~Gran}                          \affiliation{\UMD}
\author{D.A.~Harris}                      \affiliation{\york}  \affiliation{\FNAL}
\author{A.L.~Hart}                        \affiliation{\qmul}
\author{J.~Kleykamp}\thanks{\kleykampThanks}  \affiliation{\Rochester}
\author{A.~Klustov\'{a}}                  \affiliation{\ICL}
\author{D.~Last}                          \affiliation{\Rochester}  \affiliation{\upenn}
\author{A.~Lozano}\thanks{\adrianThanks}  \affiliation{\CBPF}
\author{X.-G.~Lu}                         \affiliation{\warwick}  \affiliation{\oxford}
\author{S.~Manly}                         \affiliation{\Rochester}
\author{W.A.~Mann}                        \affiliation{\Tufts}
\author{K.S.~McFarland}                   \affiliation{\Rochester}
\author{O.~Moreno}                        \affiliation{\WM}  \affiliation{\Guanajuato}
\author{J.K.~Nelson}                      \affiliation{\WM}
\author{V.~Paolone}                       \affiliation{\Pittsburgh}
\author{G.N.~Perdue}                      \affiliation{\FNAL}  \affiliation{\Rochester}
\author{C.~Pernas}                        \affiliation{\WM}
\author{M.A.~Ram\'{i}rez}                 \affiliation{\upenn}  \affiliation{\Guanajuato}
\author{N.~Roy}                           \affiliation{\york}
\author{D.~Ruterbories}                   \affiliation{\Rochester}
\author{H.~Schellman}                     \affiliation{\OregonState}
\author{C.~J.~Solano~Salinas}             \affiliation{\UNI}
\author{D.~S.~Correia}                    \affiliation{\CBPF}
\author{M.~Sultana}                       \affiliation{\Rochester}
\author{N.H.~Vaughan}                     \affiliation{\OregonState}
\author{A.V.~Waldron}                     \affiliation{\qmul}  \affiliation{\ICL}
\author{B.~Yaeggy}\thanks{\byaeggyThanks}  \affiliation{\USM}
\author{L.~Zazueta}\thanks{\lazazuetareyesThanks}  \affiliation{\WM}

\collaboration{The \minerva~Collaboration}
\noaffiliation
\date{\today}

\begin{abstract}

This paper illustrates a generic method for multi-dimensional 
reweighting of ${\cal O}(1)$~GeV neutrino interaction Monte Carlo samples. 
The reweighting is based on a Boosted Decision Tree algorithm trained on high-dimensional space in detector final-state observables.  This enables one generator's events to be reweighted so that its reconstructed particle content and kinematics distributions, as well as detector efficiency, match those of a target model.
The approach establishes an efficient way to reuse legacy Monte Carlo data, avoiding re-generation.
As an example, we test its use in a measurement of transverse kinematic imbalance of the $\mu^-$ and proton in charged-current quasielastic like $\nu_\mu$ events from the MINERvA experiment.

\end{abstract}

\maketitle

\section{Introduction and Motivation}\label{sec:introduction}

A typical analysis to estimate neutrino interaction cross sections at $\sim 1$~GeV neutrino energies uses a detector simulation to measure the efficiency to reconstruct events, to leverage data constraints to predict backgrounds, and to assess the effects of detector resolution.
The full Monte Carlo (MC) simulations begin with a generator, such as GENIE \cite{Andreopoulos:2009rq, Andreopoulos:2015wxa,GENIE:2022qrc}, NEUT \cite{Hayato_2021}, NuWro \cite{GOLAN2012499}, ACHILLES \cite{ACHILLES:PhysRevD.107.033007}, or GiBUU \cite{BUSS20121} to predict the particle content of a given neutrino interaction, and then simulate the response of the detectors to the particles. Typically, the second step, focused on the detector response, is far more computationally expensive than the first step of simulating the neutrino interaction. Even beyond this, regenerating the full MC simulation to assess differences in assumptions among different generators, or between their various versions, is inefficient due to the parameter estimation and random sampling process \cite{Pickering_2017}.

A method that would allow already simulated predictions to be reweighted to those of another generator would be computationally efficient, avoid the random sampling problem, and would allow experiments who can no longer run their detector simulation due to aging software, such as the MINERvA \cite{Aliaga:2013uqz} experiment, to test the results of new simulations.  However, a practical difficulty is that there are many degrees of freedom in neutrino interactions kinematically and they can produce from zero to many pions and knockout nucleons in their interactions in nuclei. In the extreme case, one generator can populate a specific final-state topology that is absent from another generator.
This results in events having non-overlapping particle content and kinematics in different neutrino generators.

This problem can be managed by not forcing identical particle content, but rather by forcing observable detector quantities to be identical.  For example, stable particles whose momenta are below Cerenkov threshold are effectively invisible in a water Cerenkov detector, so there would be no need for an original ``source" simulation and a desired ``target" simulation to match each other in predicting those particles.  But it would be important for simulations to agree on the numbers and momenta of $\pi^+$ that are produced, since those can be detected with high efficiency through the decay chain $\pi^+\to \mu^+\nu_\mu$, $\mu^+\to e^+\overline{\nu}_\mu\nu_e$.  Limiting the numbers of visible kinematic quantities to match detector capabilities vastly reduces the dimensionality of the problem, at the expense of making the weighting often not directly related to the models and parameters in the simulation.  For example, an event with a $\pi^+$ produced by a baryon resonance decay might be (partly) simulated by an increased weight to events where the $\pi^+$ is produced in a final-state interaction of a single nucleon knockout event.

This paper provides an example of such a practical, lower dimensional, detector-focused reweighting and its application to a measurement of transverse kinematic imbalance of $\mu^-$-proton events in the MINERvA experiment. The measurement focuses on \numuC 
charged-current quasielastic like (CCQE-like) events, where the event final-state content has one lepton accompanied with few knock-out nucleons and no meson.
The ``source'' sample generated by MINERvA using GENIE v2.12.6 (v2) is reweighted into a ``target'' sample (with the same preset) generated by GENIE v3.04.00 (v3) AR23\_20i\_00\_000 (AR23) tune, the latter of which now is the widely used GENIE version and tuning in the neutrino community \cite{Munteanu:2024, AR23notes:2025} (Technical note by DUNE collaborators on AR23 tune in preparation).
After reweighting, the kinematic variable distributions of source sample match that of the target sample, where the Kolmogorov–Smirnov test score is improved closer to zero in all trained variables and untrained but correlated variables.

The determination of weights even in this lower dimensional space benefits greatly from machine-learning techniques, in this case the Boosted Decision Tree (BDT), a tree-based method for data classification or regression. In High-Energy Physics, a BDT is a widely used multivariate technique, either as event classifier or as reweighter \cite{Yann:2013}.
In this study, we use the BDT reweighter algorithm developed by Rogozhnikov {\it et al} \cite{alex_rogozhnikov_2023_7717895, Rogozhnikov:2016}.

Elsewhere in the literature, this BDT reweighting algorithm is employed by DUNE in evaluation of oscillation parameter biases from alternative neutrino interaction models \cite{instruments5040031}. One part of the study shows that the reweighted model predicts measurable quantities, distributions of measured $Q^2$ separated by pion multiplicity, and another part shows how a movable off-axis detector can be used to correct for the bias. 
MicroBooNE's data driven model validation study \cite{PhysRevD.111.092010} examined the latter by reweighting a model to match simulated data in inclusive CC $\nu_\mu$ interactions from another prediction, where the reweighting is based on muon kinematics and true energy transfer from the neutrino to the muon. Both DUNE and MicroBooNE's studies emphasize the poor true-to-reco energy mapping despite the good agreement in trained reweight variables.
Our work has a different focus, which is demonstrating a method to reproduce directly measurable quantities; neutrino energy is not directly measurable.

A different implementation of a similar reweighting strategy was used by NOvA in the results of Ref.~\cite{nova:2021nfi} to adjust the behavior of their final-state interaction model (FSI) to reproduce different final-state hadron multiplicities.  The variables that were tuned with the BDT reweighting were limited to this specific purpose: the true $E_{\rm\scriptstyle lepton}$ and the deposited energy from protons, $\pi^\pm$, and $\pi^0$.
In contrast, our study applies the BDT reweighting in high-dimensional spaces of final-state truth quantities, aiming to transform one generator’s event sample so that its reconstructed  distributions will match those of a target model by modifying the distributions of true variables.

\section{Boosted Decision Tree Reweighter}

The goal of the weighting process is to find the right multiplier for each MC event so that,
\begin{equation}
    \text{multiplier}_{\text{bin}}=\frac{w_{\text{bin,target}}}{w_{\text{bin,source}}},
\end{equation}
where $w$ is the weight, and ``bin" represents a single region in a multi-dimensional histogram of events \cite{Rogozhnikov:2016}.  If the number of dimensions is one or two, this ratio can be calculated easily by a direct comparison between regular histograms (1D or 2D). However, it becomes impractical for higher dimensions, such as in the application here to neutrino event generation.
The CCQE-like \numuC events for example can have 1 $\mu^-$ and multiple nucleons in the final-state, whose 3-momenta form a high-dimensional set of independent variables.

\begin{figure}
    \centering
    \includegraphics[width=1.0\linewidth]{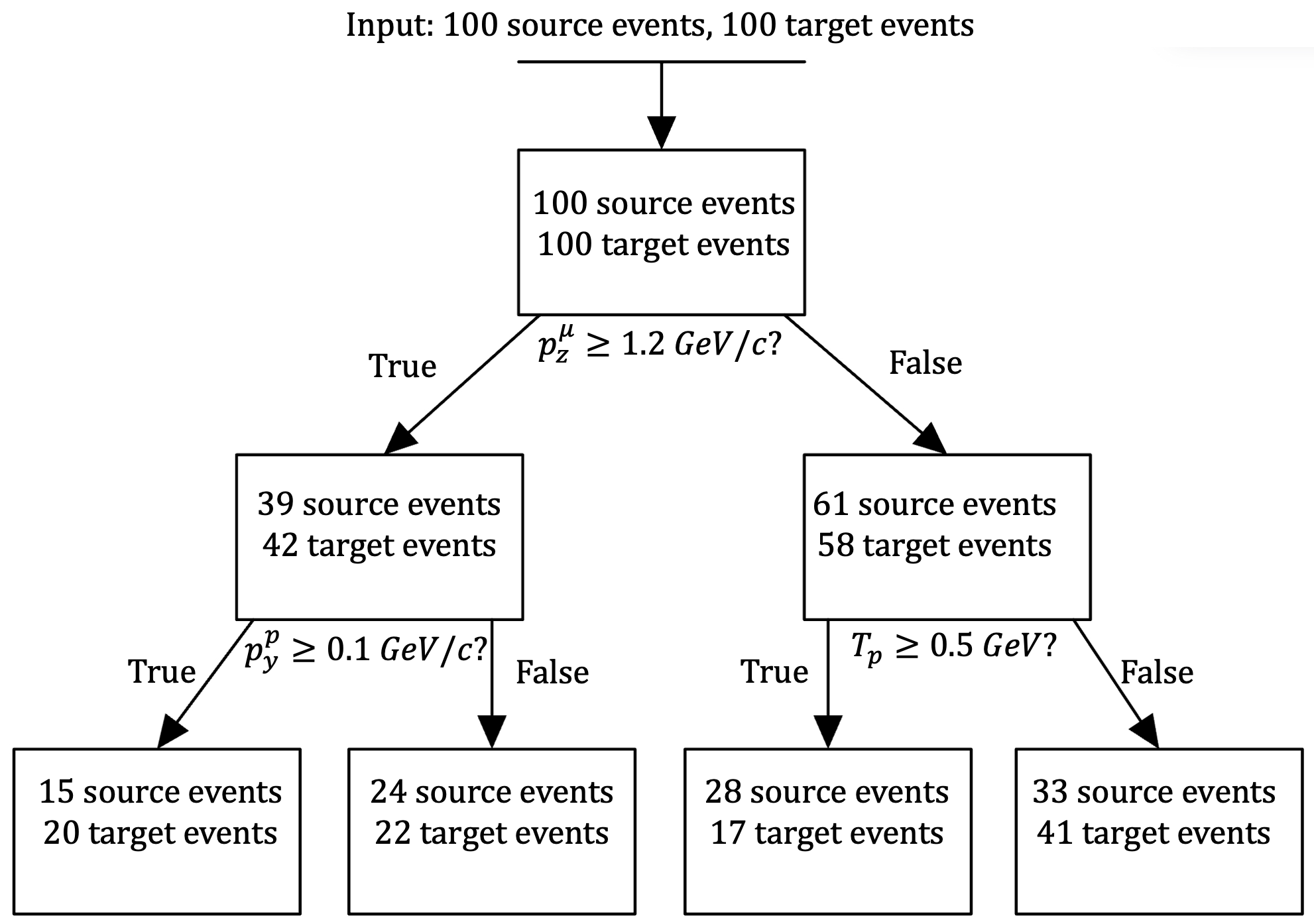}
    \caption{Example of a decision tree splitting 100 source events and 100 target events into different kinematic regions based on boolean conditions on parameters $p_z^\mu,p_y^p,$ and $T_p$.}
    \label{fig:decision-tree}
\end{figure}

In this situation, the problem becomes one of identifying which regions of the multi-dimensional space are to be reweighted to get the best agreement between the weighted source distribution and the target.
The BDT reweighter is designed to solve reweighting in a machine learning approach analogous to gradient boosting \cite{Friedman:2001}.
Data sets of the same variable spaces from source and target samples are prepared for training, which is an iterative process of building decision trees in sequence.
During each iteration, a new tree is created to recursively split the source and target events into different regions at terminal nodes, also known as the leaves of the tree structure \cite{james2023introduction,Yann:2013}, as illustrated in Figure \ref{fig:decision-tree}.
The splitting aims to maximize the symmetrized $\chi^2$ defined as
\begin{equation}\label{eq:chi2}
    \chi^2=\sum_{\text{leaf}}\frac{(w_{\text{leaf,source}}-w_{\text{leaf,target}})^2}{w_{\text{leaf,source}}+w_{\text{leaf,target}}},
\end{equation}
where $w_{\text{leaf,source/target}}$ is the sum of weights of source / target events assigned to the leaf.  
$\chi^2$ measures how much a region of source distribution is different from that of the target distribution. 
By maximizing $\chi^2$, the decision tree partitions the data into regions that are more different, therefore more relevant to reweight.
At each leaf of the tree, a prediction $\lambda_{\text{leaf}}$\, using $w_{\text{leaf,source/target}}$, is made:
\begin{equation}\label{eq:leaf_pred}
    \lambda_{\text{leaf}}=\ln \left(\frac{w_{\text{leaf,target}}}{w_{\text{leaf,source}}}\right).
\end{equation}
The source event with weight $w_{\text{event,source}}$ assigned to this leaf will be reweighted,
\begin{equation}\label{eq:event_reweight}
\begin{split}
    & w_{\text{event,source}}\to \\ 
    & w'_{\text{event,source}}=w_{\text{event,source}}\times \exp(\lambda_{\text{leaf}}),\\
\end{split}
\end{equation}
which completes the training of one tree. 
Since decision trees can easily overtrain, controlling the tree's maximum depth or pruning subtrees (branches) that are deemed too specific to training sample can help avoid such effects \cite{Yann:2013}.

After looping through many trees, the source distributions are reweighted gradually to agree with the target distribution. 
By evaluating inputs of the same variable spaces through the trees' decision chains, the reweighter is able to estimate weights for statistically independent events generated by the source generator:
\begin{equation}
    \begin{split}
        & w_{\text{event,source}}\to \\
        & w'_{\text{event,source}} 
        =w_{\text{event,source}}\times \exp\left(\sum_{\text{tree}} \lambda_{\text{tree}}\right),
    \end{split}
\end{equation}
where $\lambda_{\text{tree}}$ equals the prediction of a leaf containing this event, $\lambda_{\text{leaf}}$.

\section{Event Categorization}\label{sec:event-categorization}
In order to apply a reweighting scheme in a fixed lower dimensional space, where a few variables will be selected from the plentiful kinematic variables in neutrino MC event final-states, event categories based on final-state topologies are introduced. 
In the following subsections, a reaction plane coordinate system is introduced to describe the physics picture, and a practical approach to define event topologies and choose reweight variables for \minerva\ CCQE-like events is presented.

\subsection{Reaction Plane Kinematics}\label{sec:reaction-plane}
\begin{figure}[tp]
    \centering
    \includegraphics[width=0.9\linewidth]{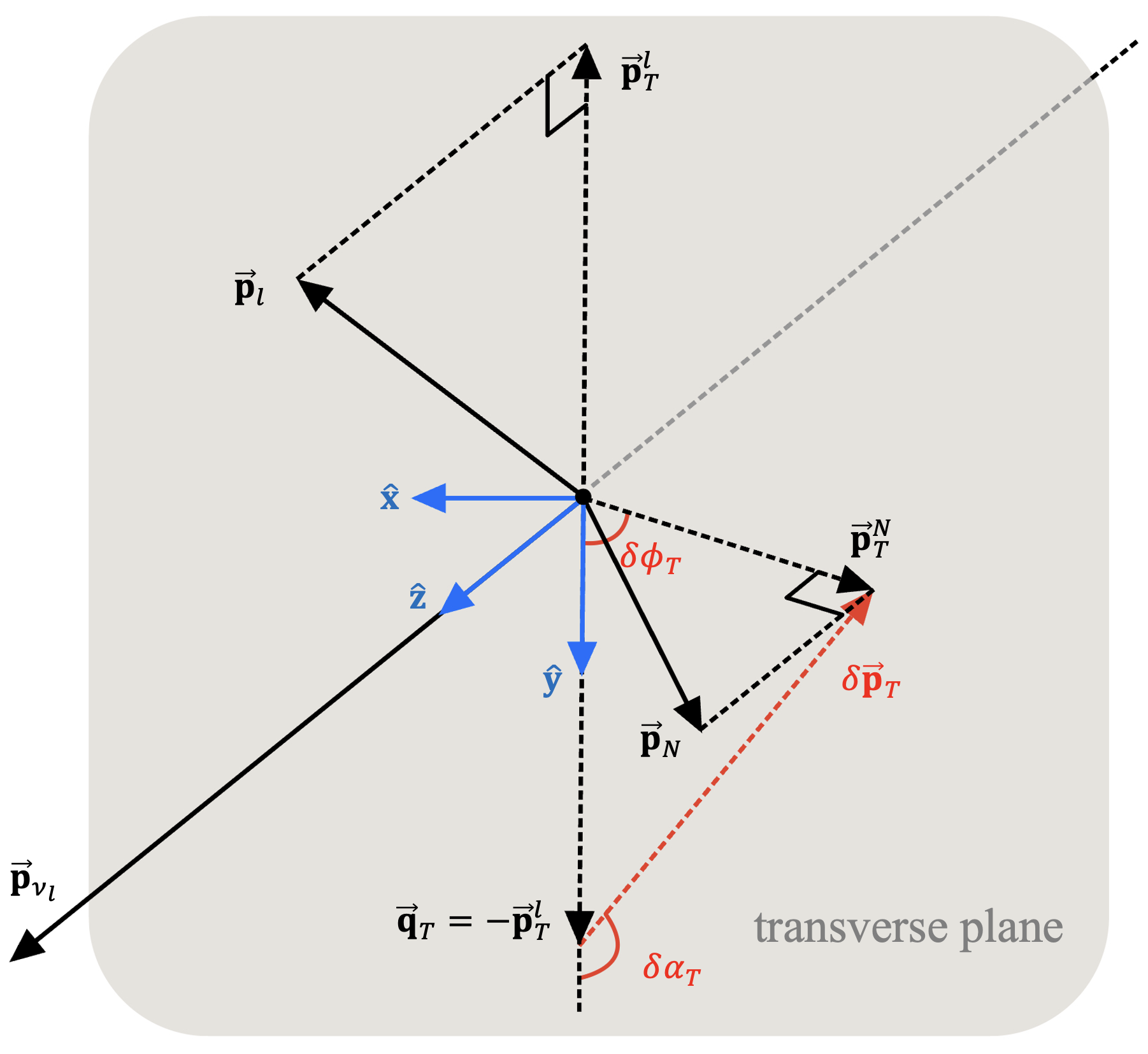}
    \caption{Schematic illustration of the single-transverse kinematic imbalance — $\delta \phi_T$, $\delta p_T$, and $\delta \alpha_T$ — defined in the plane transverse to the neutrino direction (figure derived from Figure~2 of reference \cite{Lu:2015tcr} and Figure~2 of reference \cite{Cai:2020PRD}).
    The neutrino direction and $\hat{z}$-axis is out of the page while the transverse $\hat{x}$ $\hat{y}$ is in the plane of the page.
    If the hadron momenta $\vec{p}_N$ (possibly a single particle) is observed, the TKI variables account for the differences between its transverse component and the true transverse momentum transfer $\vec{q}_T$.
    }
    \label{fig:reaction-frame}
\end{figure}

In this study, the neutrino event kinematics are defined in a reference frame in which the struck nuclei have zero initial momentum.
A reaction plane is defined by identifying the incoming neutrino direction with the $\hat{z}$-axis and defining the shared plane in which both the incoming neutrino and outgoing lepton three momenta to lie in the $\hat{y}$-$\hat{z}$ plane.
This reaction plane is for the hadron system, so the transverse lepton momentum direction is $-\hat{y}$ and the three-momentum transfer direction is $+\hat{y}$, as shown in Figure~\ref{fig:reaction-frame}. This can be done without loss of generality because the reaction is invariant under rotation around the $\hat{z}$-axis, the neutrino beam direction.

The reaction plane is instructive for understanding the kinematics of lepton and knock-out nucleon in CCQE-like processes. A transverse kinematic imbalance (TKI) will appear in any nucleon momentum in the $\hat{x}$ direction and in any difference between the muon $p_y$ and the negative of the nucleon $p_y$.
TKI effects are caused by Fermi-motion, rescattering as hadrons leave the nucleus, and missing the momentum carried by neutrons, in addition to resolution effects, such as the MINERvA studies in \cite{Lu:2015tcr, Cai:2020PRD}.
The TKI variables ($\delta \phi_T$, $\delta p_T$, and $\delta \alpha_T$) can also be built and visualized within the reaction plane in Figure \ref{fig:reaction-frame}.

\subsection{Particle Content and Detection Threshold}
Event categorization is based on the visible particle content of the event.  Only particles which are plausibly reconstructible by the detector are individually identified in these categories, and the kinematics of those particles may then enter into the training. Particles which are below detection threshold are not individually identified, and their kinematics are only considered in aggregate, if at all.  This is done to reduce the high dimension of event final-state particle information and make the machine-learning training practical.

A particle is detectable if its kinetic energy (KE) exceeds its corresponding detection thresholds. 
These are, in turn, defined by the detector design and event reconstruction methods.
In the \minerva~detector for example, protons may be tracked with reasonable efficiency from the interaction vertex only if they have a KE above $50$~MeV.
The efficiency for tracking a proton from the interaction point becomes non-zero around 50 MeV.  Below that threshold, protons are only observed calorimetrically and the proton kinetic energies below threshold are a proxy for that detection.
Neutrons (from the neutrino interaction or from secondary interactions) with kinetic energy above 10 MeV have a chance to be detected when they elastic scatter from hydrogen nuclei in the detector, when they knockout protons from carbon nuclei, and when those carbon nuclei deexcite producing photons or nucleons.

Names of, and above-threshold particles in, the categories used in this reweighter application are listed in Table \ref{tab:categories-variables}'s left and middle column.  In CCQE-like interactions of neutrinos, events with multiple neutrons over threshold are rare, and the efficiency to reconstruct a neutron increases significantly with neutron kinetic energy. Therefore as a simplification, all topologies with one or more neutrons are considered together and only the most energetic (or ``leading") neutron's kinematics are included in the reweighting scheme.  Similarly, the numbers of events with more than two protons above reconstruction threshold is small, and so such events are lumped into a single category, again with training based on the leading protons.  In this way, the kinematic information of these final-state particles can be selected from the numerous degrees of freedom in neutrino interaction. They form a lower-dimensional parameter space that is useful for detector-focused interpretation and can be more easily reproduced by  the machine learning based reweighting. 

Reweighting takes place within these categories, with each category having its own BDT reweighter trained independently of the others and used to estimate weights for events of that category exclusively.

\subsection{Reweight Variables}

\begin{table*}[t]
\centering
\setlength{\tabcolsep}{4pt}      
\renewcommand{\arraystretch}{1.1}

\resizebox{\textwidth}{!}{%
\begin{tabular}{lll}
\hline
\textbf{Topology} & \textbf{Above-threshold nucleons} & \textbf{Reweight variables} \\
\hline
0p0n  & 0 proton, 0 neutron  &
$p_y, p_z$ of muon; $\sum p_x, \sum p_y, \sum p_z, \sum T_p$ over all protons \\

0pNn  & 0 proton, $N\!\ge\!1$ neutron(s) &
$p_y, p_z$ of muon; $p_x, p_y, p_z$ of leading neutron;
$\sum p_x, \sum p_y, \sum p_z, \sum T_p$ over all protons \\

1p0n  & 1 proton, 0 neutron &
$p_y, p_z$ of muon; $p_x, p_y, p_z$ of leading proton;
$\sum T_p$ over all protons \\

1pNn  & 1 proton, $N\!\ge\!1$ neutron(s) &
$p_y, p_z$ of muon; $p_x, p_y, p_z$ of leading proton and neutron;
$\sum T_p$ over all protons \\

2p0n  & 2 protons, 0 neutron &
$p_y, p_z$ of muon; $p_x, p_y, p_z$ of two above-threshold protons;
$\sum T_p$ over all protons \\

2pNn  & 2 protons, $N\!\ge\!1$ neutron(s) &
$p_y, p_z$ of muon; $p_x, p_y, p_z$ of two above-threshold protons and
the leading neutron; $\sum T_p$ over all protons \\

others & 3 or more protons &
$p_y, p_z$ of muon; $p_x, p_y, p_z$ of leading proton;
$\sum T_p$ over all protons \\
\hline
\end{tabular}
}
\caption{\raggedright
The topologies and reweight variables for CCQE-like \numuC events.
Left column: topology names. Middle column: above-threshold final-state protons/neutrons
based on whether their kinetic energy exceeds the detection threshold.
For \minerva~detector, thresholds are 50\,MeV (proton) and 10\,MeV (neutron).
Right column: variables used for training and weight estimation.
}
\label{tab:categories-variables}
\end{table*}

Reweight variables are the parameters chosen to represent the defining final-state features in a detector. They are fed to reweighters for training and later for weight estimation. To capture the defining features of \numuC~CCQE-like topologies in \minerva~detector, we choose the momenta and calorimetric energy of above-threshold final-state particles as reweight variables.
Always included as reweight variables are $p_y$ and $p_z$ of $\mu^-$ (lepton $p_x$ vanishes in the reaction plane coordinates, by construction) and calorimetric energy, $\sum T_p$ the sum of the kinetic energy of all protons. If there are 1 or 2 above-threshold protons, $p_x, p_y,$ and $p_z$ of these protons are also included; if there are 3 or more above-threshold protons, only $p_x, p_y,$ and $p_z$ of the leading proton are included; and if there is no above-threshold proton, the momenta $\sum p_x, \sum p_y, $ and $\sum p_z$ summed over below-threshold protons are included. 
If there are 1 or more above-threshold neutron(s), only the leading neutron's $p_x,p_y$ and $p_z$ are included. These reweight variables are summarized in Table \ref{tab:categories-variables}'s right column.  

The above choices are optimized for the goals of this study, which are to reproduce variables of interest that are reconstructable in the MINERvA detector and used for its physics results.  If the goal were to study multi-neutron production in neutrino interactions, for example, a different set of categories might be chosen.
The method could also be adopted to event types beyond CCQE-like, for example, the charged-current events with 1 pion in the final-state (CC1$\pi$), for which the reweight variables may include the momenta of above-threshold final-state pion. A general approach would: (1) categorize the events based on the final-state particle content of interest and detection thresholds, (2) choose reweight variables to reflect features seen in the detector, and (3) train reweighters within each category to determine weights for events.

\section{Reweight Results}
\begin{figure}[tp]
    \centering
    \includegraphics[width=1\linewidth]{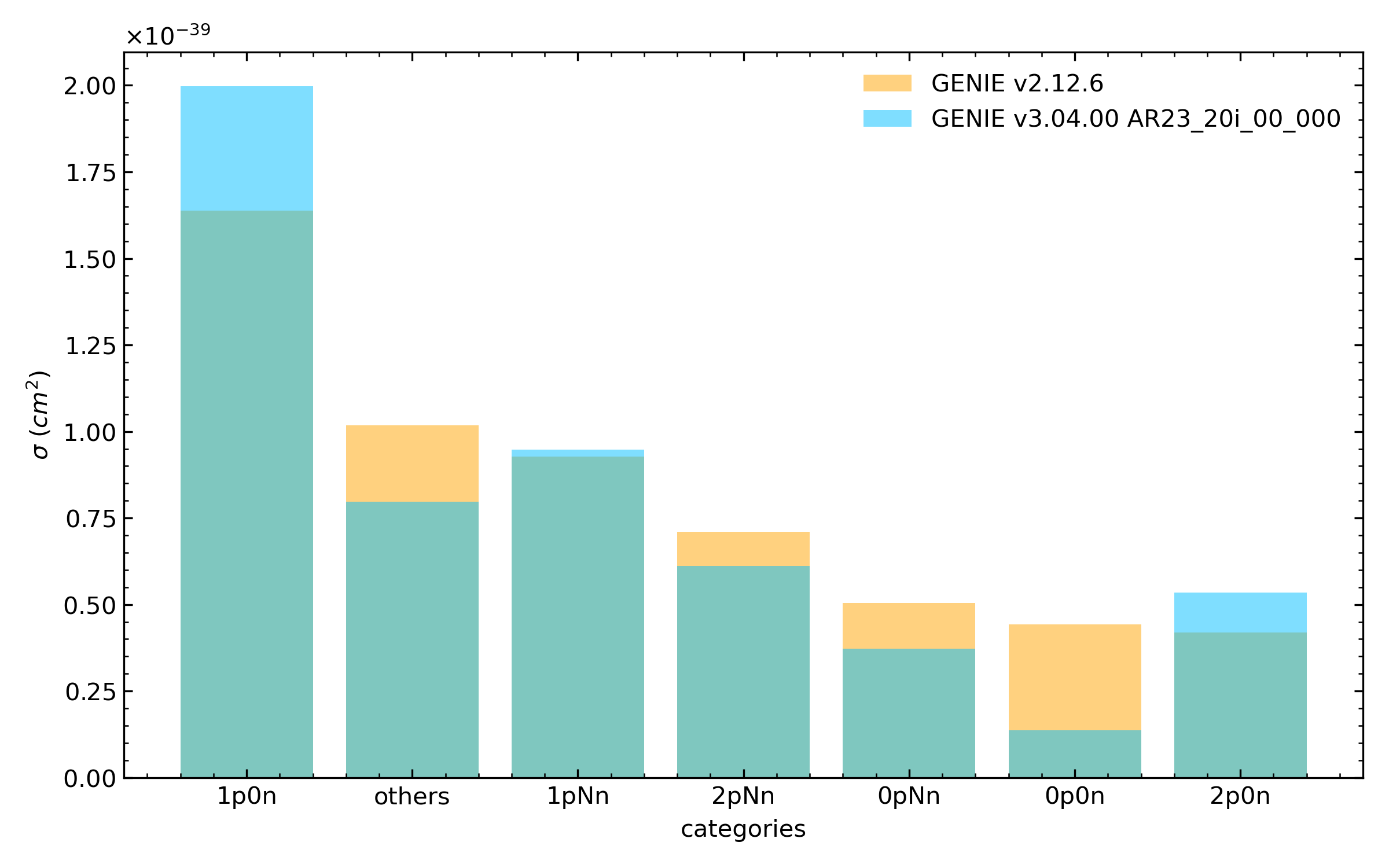}
    \caption{Categorical histogram of \minerva\ ME CCQE-like \numuC cross section contributed from the 7 categories listed in Table \ref{tab:categories-variables}. Orange: \genieTwo. Blue: \genieThree. 
    }
    \label{fig:categories}
\end{figure}

This section presents the results of reweighting \minerva~medium energy (ME) \numuC~CCQE-like events from source generator GENIE v2.12.6 to target generator GENIE v3.04.00 AR23\_20i\_00\_000.
GENIE v2 is chosen as the source because MINERvA's large sample of simulated events was generated with this version. The neutrino events are analyzed through ROOT's Python interface \cite{BRUN199781, rene_brun_2019_3895860} and NUISANCE event record format \cite{Stowell:2016jfr}. Four million event source and target samples were generated using the MINERvA ``medium energy" forward-horn current (neutrino dominant) beam \cite{Zazueta_2023, PhysRevD.94.092005}.
The GENIE v2 samples were pre-processed to remove a category of unphysical final-state interaction (FSI) events, see discussion at \cite{Harewood:2019rzy}.
However, they were not replaced with no-FSI events as is done in the usual MINERvA analyses; instead the BDT is deciding what should be done.
To ensure that source and target model share the same phase space, the 2p2h events with $q_3>1.2$ GeV/c in GENIE v3 are dropped, as this phase region is cut out in GENIE v2's default implementation of Valencia 2p2h model \cite{Nieves:2011pp}.\footnote{Upon request, MINERvA can supply a version of GENIE v2 for use with events generated and distributed with our open data product.  It has all the functionality of the original GENIE v2 code, plus options to turn on bug fixes and back ported 2p2h functionality from later versions of GENIE. The code and its dependencies have been modified to build on c.2025 era software platforms.}
CCQE-like events from source and target samples were selected and divided into subsamples according to the seven categories listed, and seven reweighters are trained independently within the categories using corresponding reweight variables. The reweighter architecture's number of trees, depth of trees, and minimum sample at leaves are tuned for improving agreement with target MC for each category. 
For category 1p0n, ensemble of 20 trees with maximum depth of 30 are trained. For all other categories, ensembles of  100 trees with maximum depth of 4 are trained.
(The need for a different architecture for the 1p0n sample is an empirical finding: the default architecture fails to reproduce the target sample as measured by the metrics in the following section.)
All training requires at least 30 events at each leaf.
The source and target models have different predictions for cross sections in each of the categories, as shown in Figure~\ref{fig:categories}.
Reweighters modify the shape of distributions, but will not match the total cross section of each category.
A normalization constant per category is calculated to change the total cross section of a given category from the statistically independent GENIE v2 test sample prediction to that of the target.
Categories are then combined to make a total prediction for observables.

\subsection{Reweighting Performance with Combined Categories}\label{subsec:rwt-performance-combined-categories}
\begin{figure*}[tp]
    \centering
    \includegraphics[width=1\linewidth]{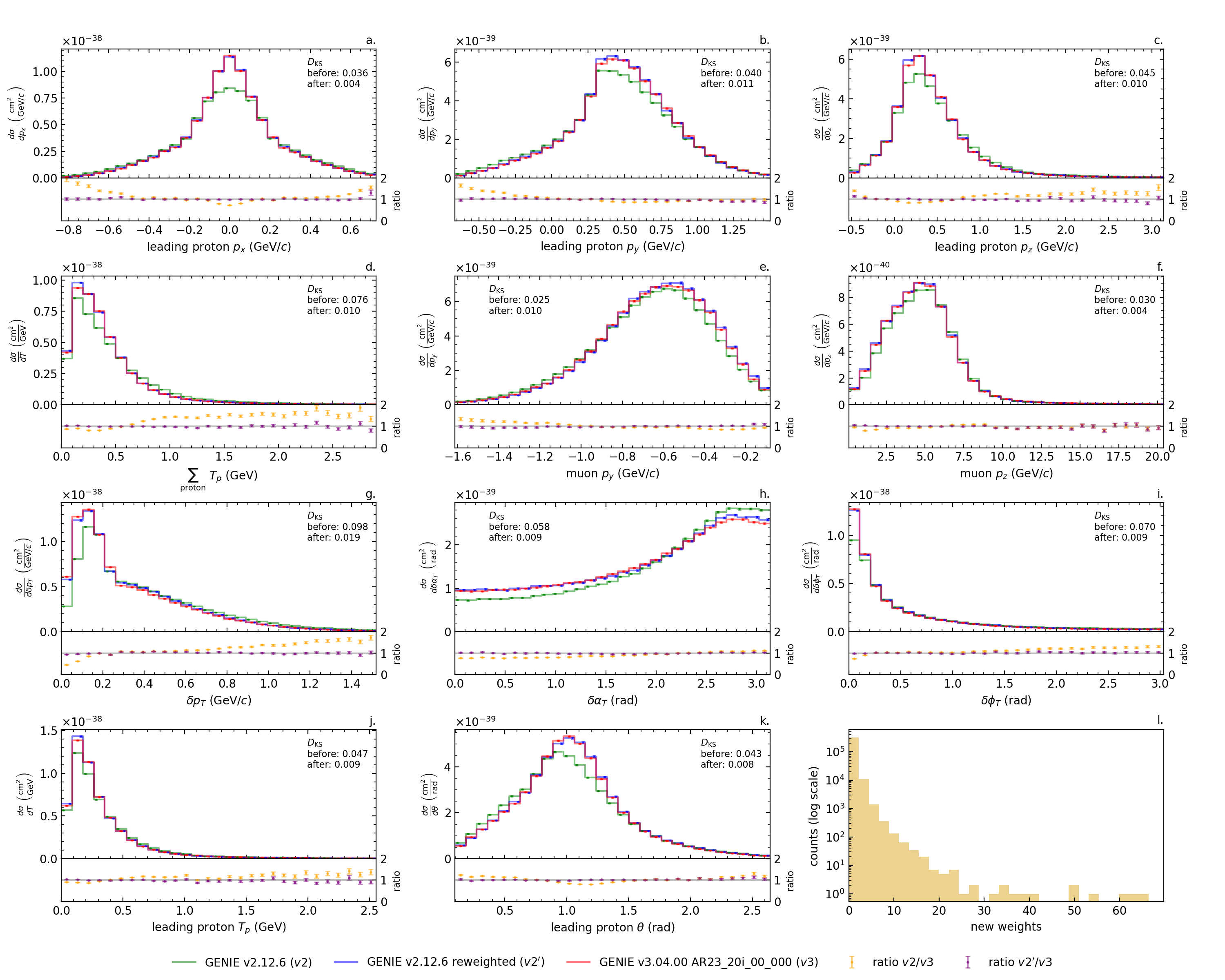}
    \caption{
        Differential cross sections of categories ``1p0n'', ``1pNn'', ``2pNn'', ``2pNn'', and ``others'' combined are plotted with respect to leading proton $p_x, p_y,p_z$ (a, b, c); calorimetric energy $\sum T_p$ (d); $\mu^-\ p_y,p_z$ (e, f); TKI variables $\delta p_T,\delta \alpha_T,\delta \phi_T$ (g, h, i); and leading proton $T_p,\theta$ (j, k).
        A frequency histogram of weights (l) is also shown.  Error bars (visible only in the ratios) are statistical only.
        Green: test sample \genieTwo~(v2). Blue: reweighted test sample (v2$'$). Red: target sample \genieThree~(v3).
        cross section ratios of v2 and v2$'$ comparing to v3 are plotted under each histogram, in yellow and purple respectively.
        K-S test statistic $D_{\text{KS}}$ before (v2 comparing to v3) and after (v2$'$ comparing to v3) reweighting is printed on each histogram.
        }
    \label{fig:combined-test}
\end{figure*}
\begin{figure*}[tp]
    \centering
    \includegraphics[width=1\linewidth]{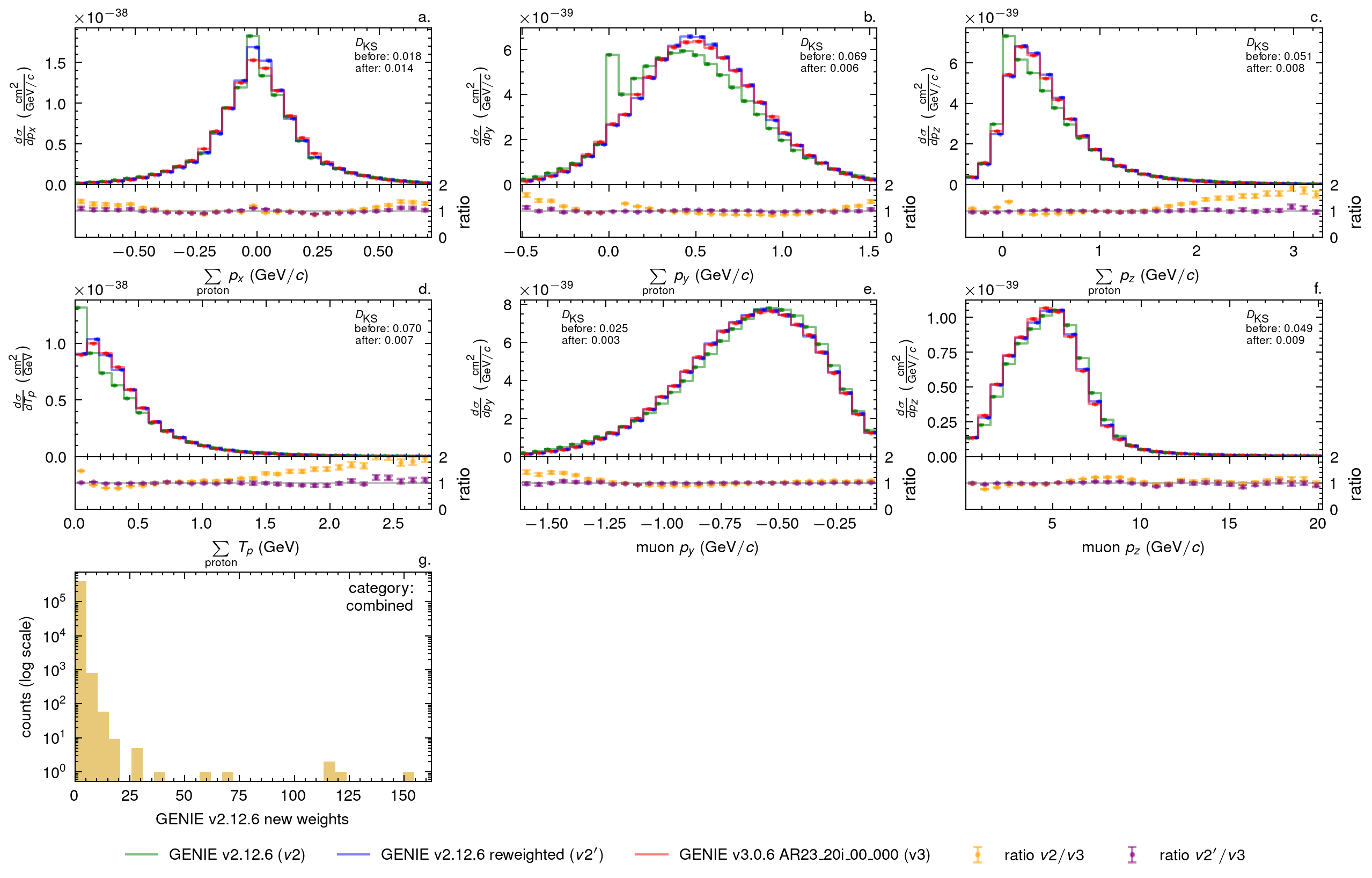}
    \caption{
        Differential cross sections of all categories combined are plotted with respect to calorimetric momenta $\sum p_x, \sum p_y, \sum p_z,$ and energy $\sum T_p$ summed over all final-state protons (a, b, c, d); and $\mu^-$ $
        p_y,p_z$ (e, f).
        A frequency histogram of weights (g) is also shown.
        Error bars (visible only in the ratios) are statistical only.
        Green: test sample \genieTwo~(v2). Blue: reweighted test sample (v2$'$). Red: target sample \genieThree~(v3).
        cross section ratios of v2 and v2$'$ comparing to v3 are plotted under each histogram, in yellow and purple respectively.
        K-S test statistic $D_{\text{KS}}$ before (v2 comparing to v3) and after (v2$'$ comparing to v3) reweighting is printed on each histogram.
    }
    \label{fig:combined-sumProton-test}
\end{figure*}
For evaluation of the performance, the resulting weights are applied to a statistically independent sample of GENIE v2.12.6 then compared to a sample from the target GENIE v3.04.00 AR23\_20i\_00\_00.
This subsection discusses the performance for the reweighted categories after they are combined. The individual results for each category can be found in Appendix \ref{appx:figures}. 

The combined differential cross sections of categories with 1 or more above-threshold proton(s) (``1p0n'', ``1pNn'', ``2p0n'', ``2pNn'', ``others'') in the final-state are shown in Figure \ref{fig:combined-test}; reweight variables (leading proton's momenta, proton calorimetric energy, and $\mu^-$ momenta) along with untrained observables such as TKI variables $\delta\alpha_T,\delta p_T, \delta \phi_T$, leading proton kinetic energy $T_p$ and angle $\theta$ are plotted.
The combined differential cross sections of all categories (including ones without protons, ``0p0n'' and ``0pNn'') are shown in Figure \ref{fig:combined-sumProton-test}. Only the trained reweight variables are plotted, though the proton calorimetric momenta and energy are the sum of separately trained calorimetric low KE and tracked high KE components.
Frequency histograms are converted to differential cross sections, where subsample cross sections of the source test sample are scaled to match that of the target sample, so the total cross sections combined are matched in magnitude.

As shown in Figure \ref{fig:combined-test},  reweighters are able to identify the differences across source and target samples, which are mostly compensated by the source sample's new weights.
The two-sample Kolmogorov-Smirnov (K-S) test statistic $D_{\text{KS}}$ is defined as the maximum distance between two empirical cumulative distribution functions $F_m(x)$ and $F_n(x)$,
\begin{equation}
    D_{\text{KS}}=\max_x|F_{m}(x)-F_{n}(x)|,
\end{equation}
where $m$ and $n$ are the samples being compared \cite{Press:2007ipz}\footnote{Ref. \cite{Melo:2009KStest_fireball} shows an example of usage in nuclear physics, where K-S test is applied to rapidity distributions to identify nuclear fragmentation processes.}.
$D_{\text{KS}}$ before and after reweighting is printed on the subplots of Figure \ref{fig:combined-test} for each kinematic variable. 
The expected distribution of the K-S test statistic under the null hypothesis, that the two samples follow the same underlying distribution, depends on the sample size and the distribution of weights in the sample. Accordingly, it is difficult to use the K-S test to assess compatibility with the null hypothesis after weighting, but we can use it to demonstrate the improvement in agreement between the two samples by the reduction of the K-S statistic.
The reweight effects are also transferred to the variables derived from $\mu^-$ and leading proton momenta, such as TKI variables, although they are not part of the training process.  This demonstrates the correlations among the target variables in the training are correctly reproduced.

Ratios of source distributions comparing to target distributions before and after reweight are plotted for reweight variables and derived quantities. In most kinematic regions, these ratios tend to 1 after reweight.
This is obtained even for the TKI variables.
In Figure~\ref{fig:combined-sumProton-test}, dramatic differences are observed in calorimetric momenta and energy, where spikes at zero are present in source test sample but not in target sample. They correspond to zero KE proton events coming from ``0p0n'' and ``0pNn'' categories (see their cross sections in Appendix \ref{appx:figures} Figure \ref{fig:0p0n-test} and \ref{fig:0pNn-test}). The reweighters remove the spikes by assigning zero weights to such events, and a large number of these zero weights can be seen in these categories in Figure~\ref{fig:0p0n-test} and \ref{fig:0pNn-test}. These zero weights reflect the reweighter's decision tree aggressively maximizing the symmetrized $\chi^2$ defined in Equation \ref{eq:chi2} to address the differences in the proton distributions.

Details about the mechanism that caused these spikes seen in GENIE v2 are reported in Appendix \ref{appx:spikes-GENIE2}. There are numerous other differences between GENIE v2 and v3, more than can be described here, and the BDT is easily able to take care of them.

\subsection{Reweight Training Using Smaller Set of Categories and Variables}\label{sec:reweight-less-complete-set}

In this subsection, the impact of choices of categories and reweight training variables is explored, specifically the case where they contain less information about the events than the default choices listed in Table \ref{tab:categories-variables}.  We look at our ability to reproduce two test measurements: muon + proton TKI variables for events with a trackable proton, and visible proton energy for all events.
In a four step study, we first try a single reweighter trained on all CCQE-like events (single category) using only muon kinematics ($p_y$, $p_z$).
Second, with the same single category, the sum of proton kinetic energy is included in training.  The sum of kinetic energy is an observable in the detector, but is not directly used in forming TKI variables. 
In the third and fourth scenarios, the single category is divided into two samples, one with no above-threshold protons and the other with one or more protons above-threshold.  Two reweighters are trained separately using muon kinematics and sum of proton kinetic energy ($\underset{\text{proton}}{\sum}T_p$).
Finally in the fourth scenario, leading proton kinematics ($p_x,p_y,p_z$) are also included in reweight training.

\begin{figure*}[tp]
    \centering
    \includegraphics[width=1\linewidth]{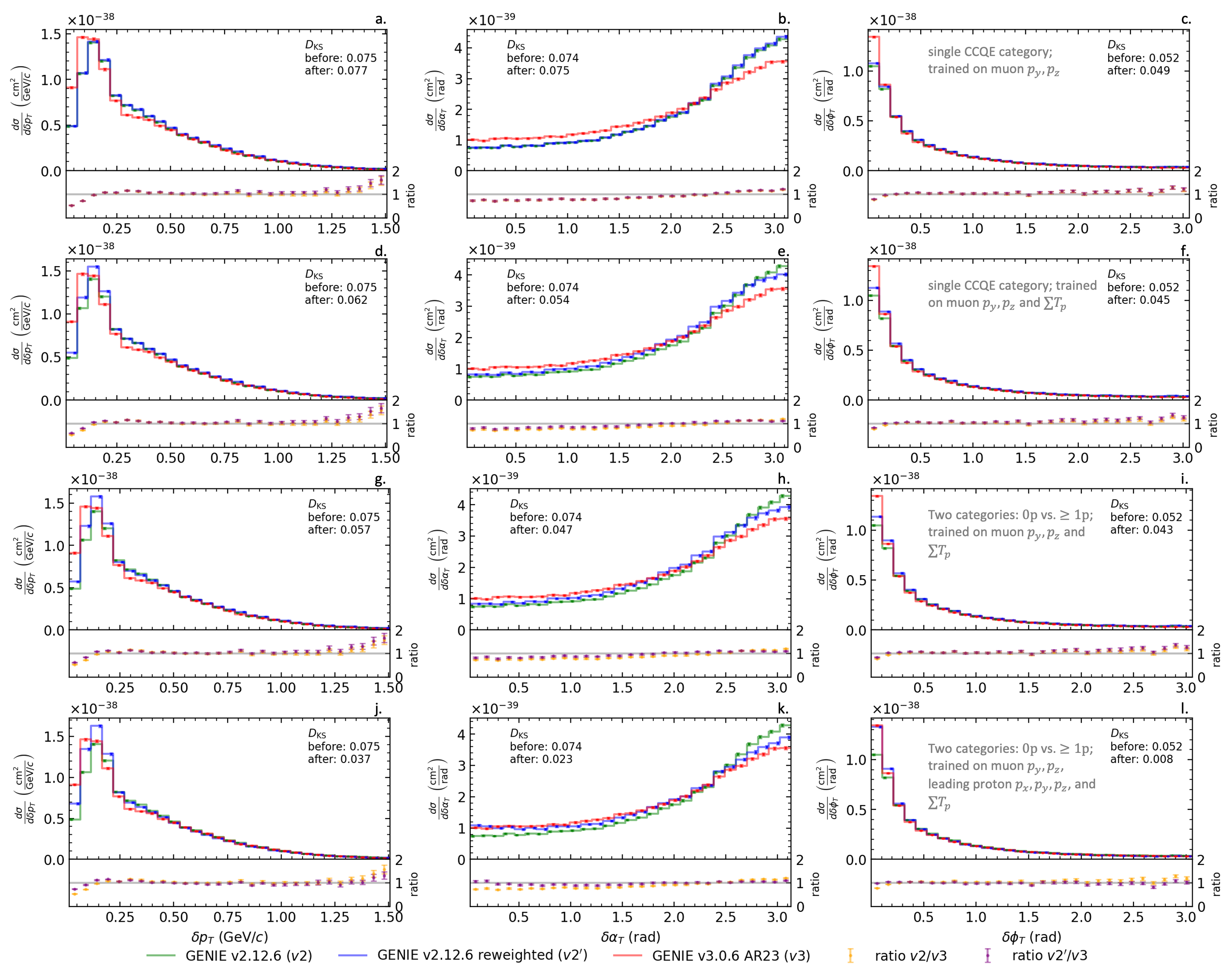}
    \caption{
    Differential cross sections with respect to TKI variables using alternative training schemes with less complete sets of training variables and categories.
    (a, b, c): single CCQE category trained on muon $p_y,p_z$. (d, e, f): single CCQE category trained on muon $p_y, p_z$, and $\underset{\text{proton}}{\sum} T_p$. (g, h, i): the single CCQE category is broken into those with no above-threshold proton and those with $\ge 1$ above-threshold protons, trained on muon $p_y, p_z$ and $\sum T_p$. (j, k, l): same categories as g, h, i, but leading proton $p_x, p_y,p_z$ are also trained in reweight.
    Only events with one or more above-threshold proton(s) are plotted for these TKI variables.
    }
    \label{fig:alternative-TKI}
\end{figure*}

The differential cross section with respect to TKI variables are plotted in Figure \ref{fig:alternative-TKI}.
As expected, no visible improvements are seen in top panel (a, b, c), since proton kinematics are never part of the reweight.
$D_{KS}$ score, which can be interpreted as the absolute value of maximal difference between two distributions, are reduced gradually as we add subsamples and additional training variables to the reweighter from top to bottom panels of Figure \ref{fig:alternative-TKI}. 
When leading proton kinematics are included in training (j, k, l), all three distributions are improved significantly, especially in low $\delta\phi_T$ region where $v2'$ almost overlaps with $v3$. 
This is a direct consequence of training on proton kinematics, because leading proton tends to travel in the opposite of muon transverse direction and events of such populate low $\delta\phi_T$ region.

This study shows that it is important to separate out the individual trackable proton category and train on the variable of interest for total visible proton energy to produce matched TKI distributions.  Note that none of the these more limited training scenarios achieve the quality of the results presented in the full division of categories in Figure \ref{fig:combined-test}, which add more subsamples with subleading protons and neutrons, and more kinematic information about these particles.
The conclusion we draw from this exploration is that it is important to validate the performance of the reweighter on the desired target distribution, particularly when it is not directly used in the training sample.

\begin{figure*}[tp]
    \centering
    \includegraphics[width=1\linewidth]{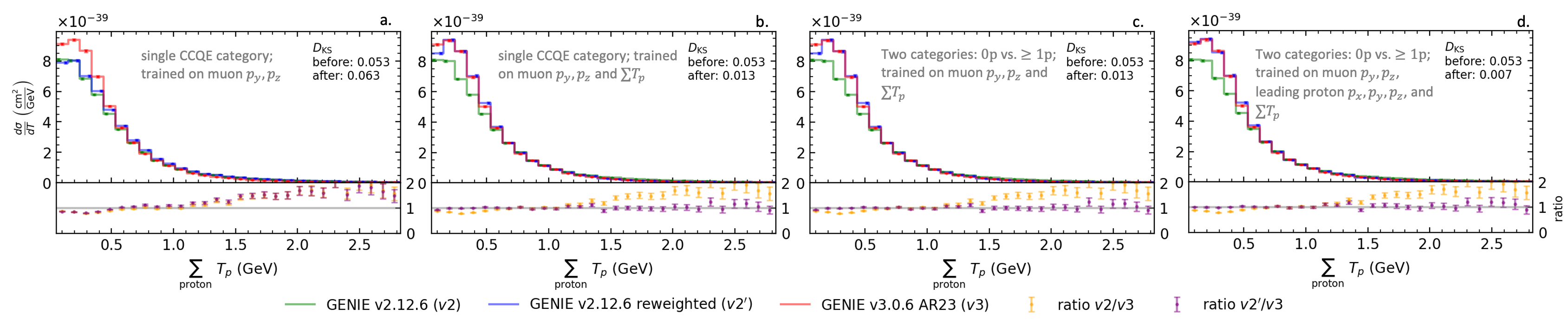}
    \caption{
    Differential cross sections with respect to $\underset{\text{proton}}{\sum} T_p$ using alternative training schemes with less complete sets of training variables and categories.
    From left to right, training variables and categories are progressively more complete as specified on subplots (a, b, c, d).
    }
    \label{fig:alternative-sumTp}
\end{figure*}
\subsection{Application: Efficiency Calculations for cross section Measurements}
\begin{figure}
    \centering
    \includegraphics[width=0.9\linewidth]{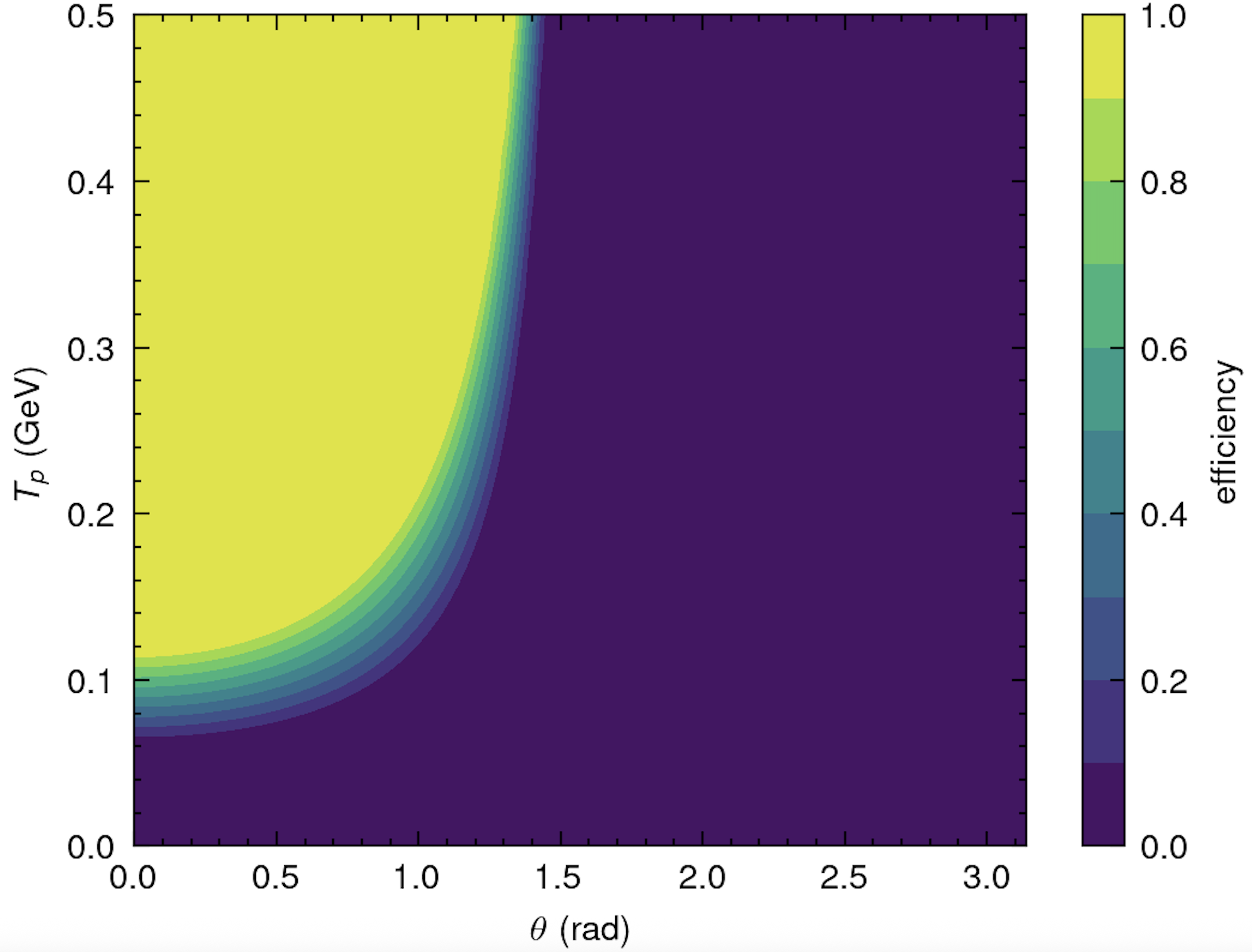}
    \caption{Contour plot of proton detecting efficiency model $\epsilon(\theta_{p},T_p)$. Efficiency ranges from 0 to 1. \minerva~has high efficiency in detecting forward traveling (low $\theta$) energetic (high $T_p$) protons.}
    \label{fig:efficiency}
\end{figure}

In this subsection, the detector efficiency of reconstructed events is measured from the test sample, the reweighted test sample, and the target sample to demonstrate the reweighter's ability to recreate the target model's prediction of detection efficiency.  Since the \minerva~detector uses planar targets and scintillator modules, it is better at measuring particles with high energy that travel along the beam axis than those which travel transverse to the beam direction. This effect must be accounted for in the proton tracking efficiency model which is an input to a cross section measurement.
Correcting for the efficiency is a standard step in producing a cross section. When efficiencies are small, the fractional uncertainties are large and almost always must be estimated with the help of a full simulation.
The correlations between measured variables of interest and the angles of produced particles with respect to the beam depend on the model, and are different between the source and target model in this study. 
In this study, we use a toy model of the efficiency, $\epsilon$, as a function of the leading proton angle with respect to the neutrino direction, $\theta_{p}$, and its kinetic energy, $T_p$.  This toy model, $\epsilon(\theta_{p},T_p)$ is given by:
\begin{equation}
    \epsilon(\theta_{p},T_p)=\min\left(\max\left(\frac{T_p\cos(\theta_{p})-60~\text{MeV}}{60~\text{MeV}}, 0.0\right),1\right),
\end{equation}
where 60 MeV was chosen above the KE selection threshold to ensure that the reweighting effectively describes the target model for events whose efficiency is affected by the threshold.
Events with high $T_p$ and low $\theta_p$
have efficiency close to one in this toy model.  The efficiency model is plotted for reference in Figure~\ref{fig:efficiency}.

To study the effect of this reweighting on the extraction of cross sections, we consider the application of this efficiency model to a two-dimensional (2D) differential cross-section with respect to the muon transverse momentum and the TKI variable $\delta p_T$ for events with one or more above-threshold protons for the MINERvA detector. The bin-wise differential cross sections, $N$, and the efficient cross section, $M$, are defined as,
\begin{equation}
    \begin{split}
        N_{ij} &= \sum_{\text{event}_k  \text{ in bin}_{ij}} C_{ij} ,\\
        M_{ij} &= \sum_{\text{event}_k  \text{ in bin}_{ij}} C_{ij}\times{\epsilon(\theta_{p,k},T_{p,k})},\\
    \end{split}  
\end{equation}
where $k$ is the index over simulated events, $i$ and $j$ index the bins in $\delta p_T$ and $p^T_\mu$, respectively, and $C_{ij}$ is the conversion factor from event rate to flux averaged differential cross section. Note that the efficiency is a function of the events $k$, and not directly of $i$ and $j$, so the model provides the connection between those two.  The efficiency for the bin, the figure of merit in this study, is defined as their ratio,
\begin{equation}
    \phi_{ij} = M_{ij}/N_{ij}.
\end{equation}
In a real experiment, $M_{ij}$ would be a measurable quantity, an event rate corrected for detector smearing, related to the true correlation between the bins of the differential cross section, and the model-dependent determination of $\phi_{ij}$ is used to extract the cross section $M_{ij}$,  with systematic uncertainties to represent uncertainties in the model's calculation of the efficiency.

\begin{figure*}[tp]
    \centering
    \includegraphics[width=1\linewidth]{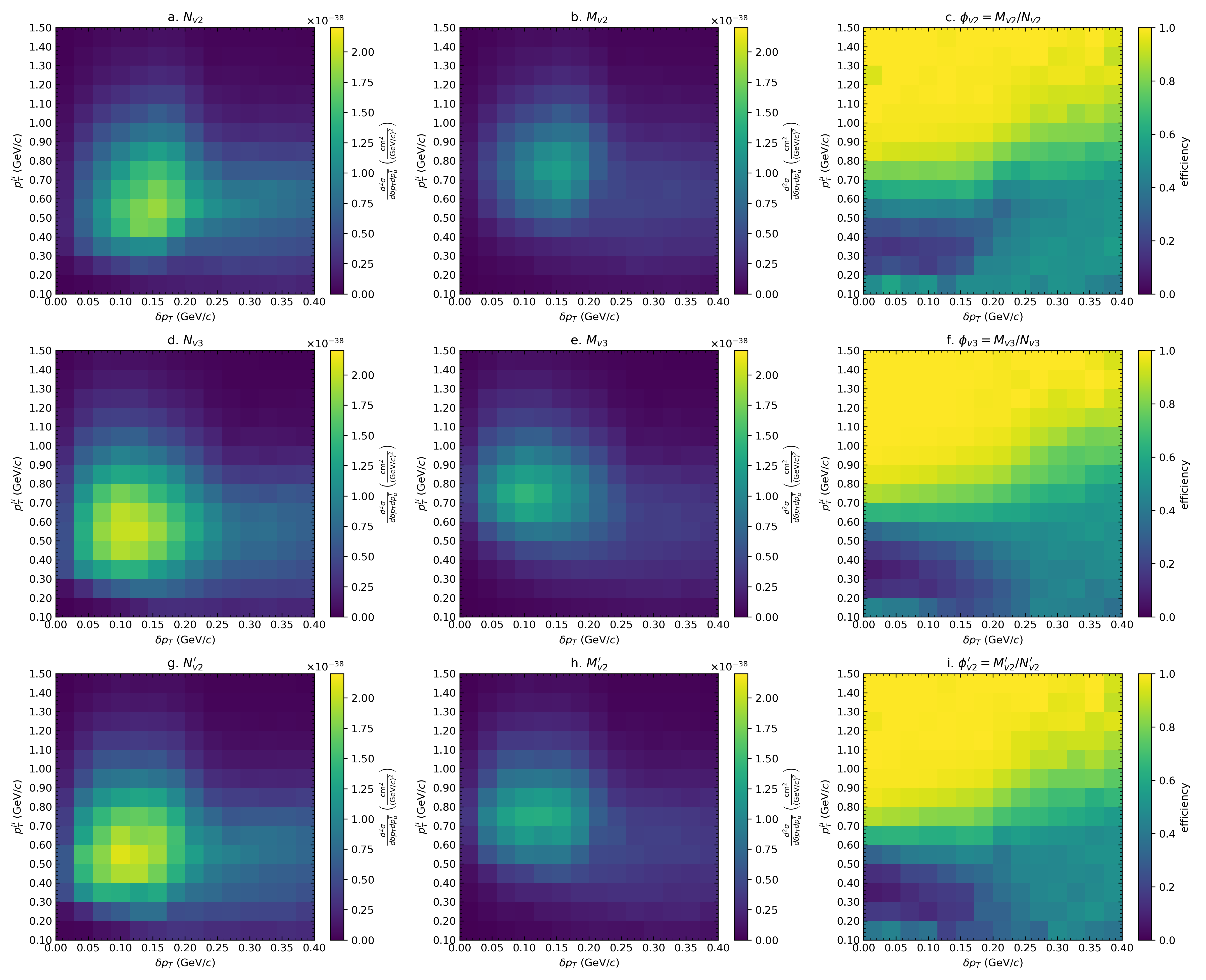}
    \caption{2D differential cross sections with respect to $\delta p_T$ and $p^T_\mu$ and efficiency of bins. 
    Top panel: true cross section $N$, efficient cross section $M$, and efficiency $\phi$ of test sample \genieTwo~(v2). 
    Middle panel: the same quantities for target sample \genieThree~(v3). Bottom panel: the same quantities for \genieTwo~sample reweighted (v2$'$).
    }
    \label{fig:NMEff-test}
\end{figure*}

\begin{figure*}[tp]
    \centering
    \includegraphics[width=1\linewidth]{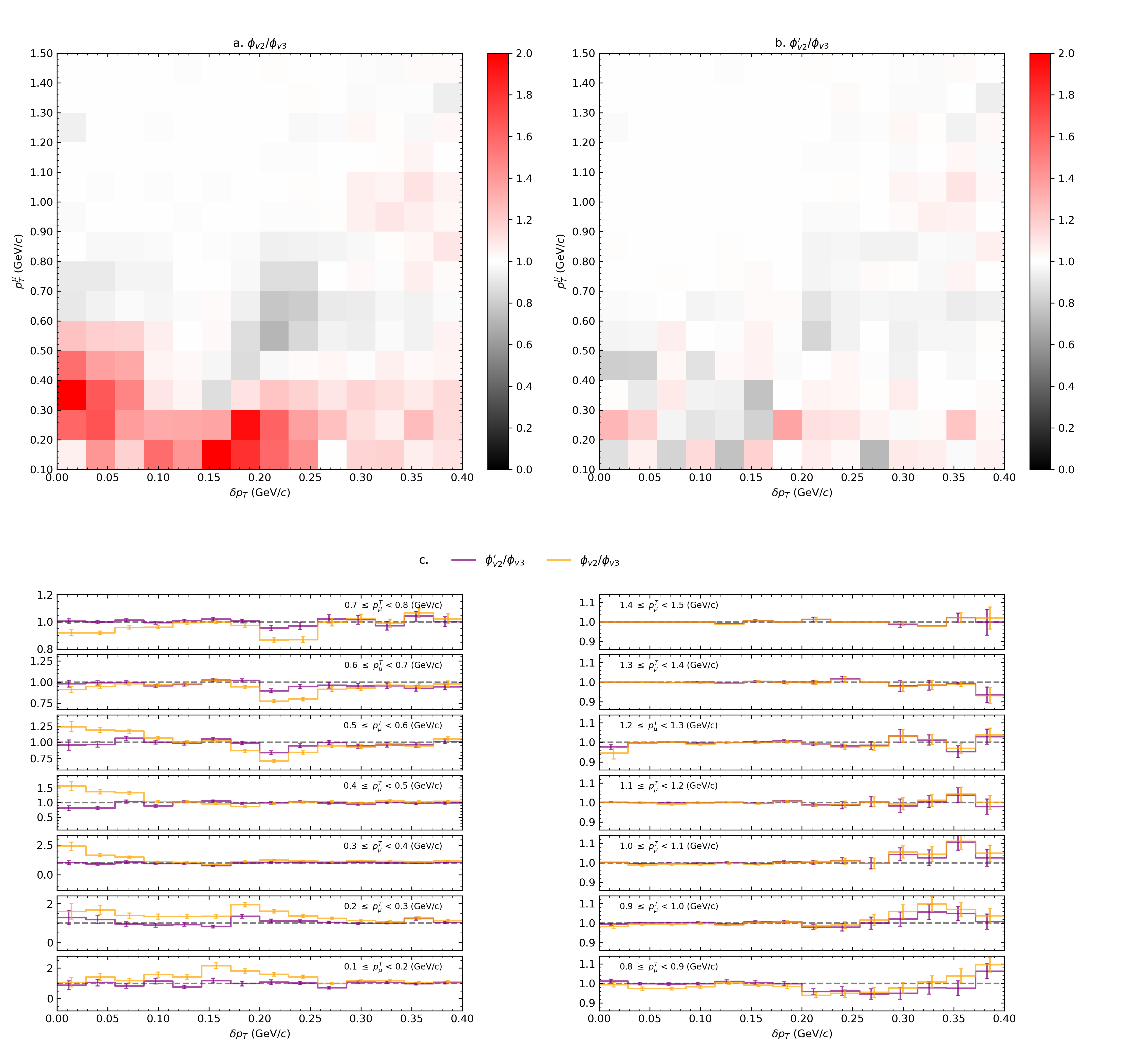}
    \caption{2D efficiency ratio plot. 
        (a): $\phi_{\text{v2}}/\phi_{\text{v3}}$, efficiency of source sample \genieTwo\ (v2) divided by efficiency of target sample \genieThree\ (v3). 
        (b): $\phi_{\text{v2}}'/\phi_{\text{v3}}$, efficiency of reweighted source sample (v2, primed) divided by efficiency of target sample.
        (c): The same ratios plotted as step functions in bins of $p^T_\mu$. (c)'s left panel: bins of $0.1\le p_\mu^T<0.8 ~(GeV/c)$; (c)'s right panel: bins of $0.8\le p_\mu^T<1.5 ~(GeV/c)$. $\phi_{\text{v2}}/\phi_{\text{v3}}$ is in yellow, and $\phi'_{\text{v2}}/\phi_{\text{v3}}$ is in purple.
    }
    \label{fig:Eff2Eff3-step-test}
\end{figure*}

The ``true'' differential cross section with respect to $\delta p_T$ and $p^T_\mu$, $N$, the efficiency weighted cross section, $M$, and efficiency $\phi$ extracted from test sample, reweighted test sample, and target sample are shown in Figure~\ref{fig:NMEff-test}. As can be seen, the efficiency in both models is lower at low $p^T_\mu$ and $\delta p_T$, but the dependence on those variables is different in the two models.  Figure~\ref{fig:Eff2Eff3-step-test} compares the ratio of the efficiencies between the source, GENIE v2, and target, GENIE v3 AR23, models before and after the reweighting.  The reweighting significantly reduces the difference between the two models, with some differences remaining at the lowest $p^T_\mu$ and $\delta p_T$ where the statistics for both the efficiency weighted and true samples are very low. 

\subsection{Application: Truth-to-Reconstruction Migration Unfolding and Efficiency of the Signal for an Example cross section}\label{subsec:migration-matrix}

\begin{figure*}[tp]
    \centering
    \includegraphics[width=1\linewidth]{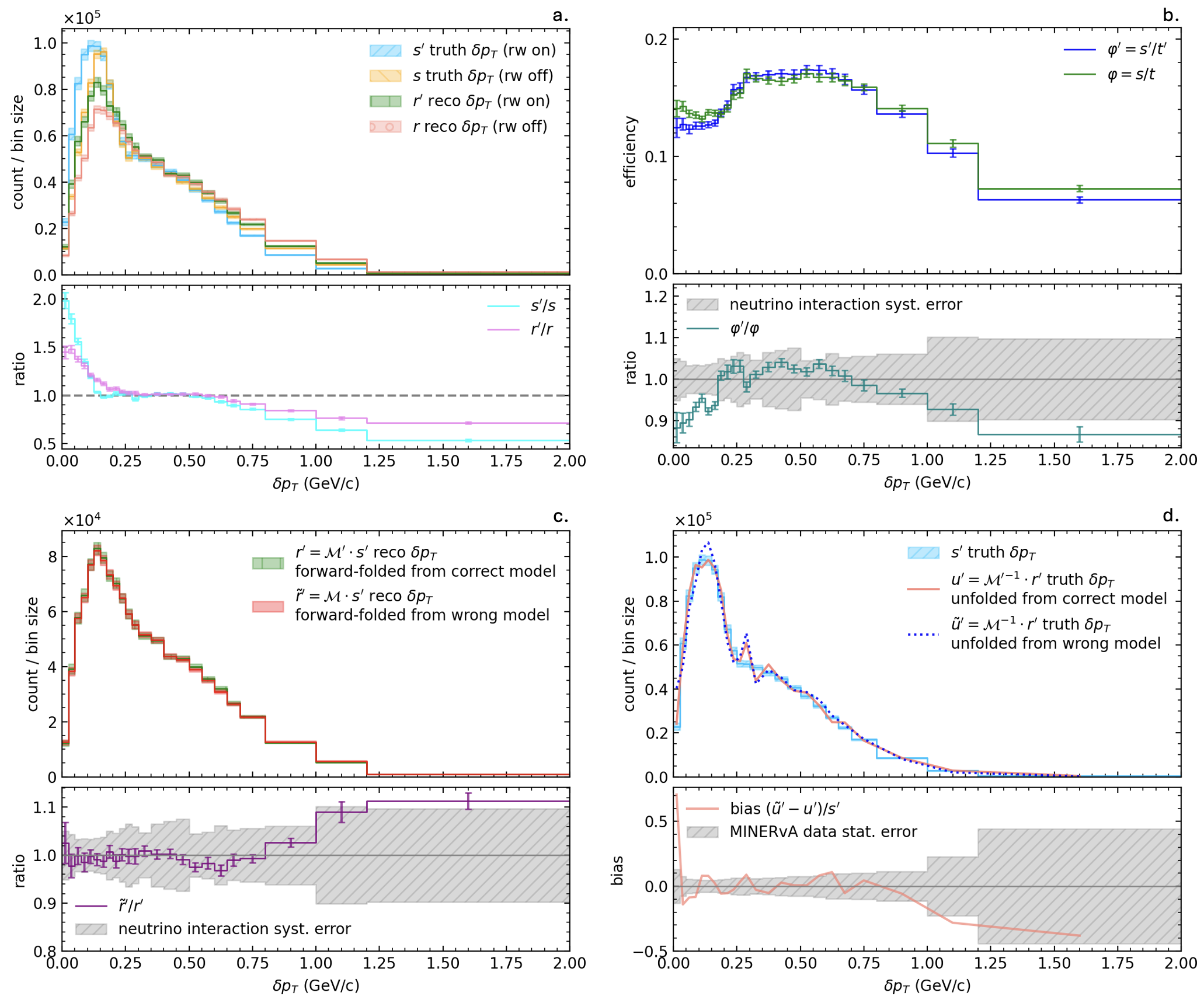}
    \caption{
        Truth and reconstructed (reco) $\delta p_T$ distributions, reco selection efficiencies, forward-folded reco $\delta p_T$ from migration matrix, and unfolded truth $\delta p_T$ from migration matrix's pseudo-inverse. Primed (unprimed) quantities are those of GENIE v2.12.6 with BDT reweight turned on (off).
        (a): selected signal events' $\delta p_T$ truth distributions, $s$ and $s'$, and reco distributions, $r$ and $r'$. 
        Ratios $s'/s$ and $r'/r$ are plotted in bottom panel.
        (b): reco selection efficiency $\varphi'$ and $\varphi$ where $t$ and $t'$ are total number of events. Ratio $\varphi'/\varphi$ is plotted in bottom panel.
        (c): $r'=\mathcal{M'}\cdot s'$ comparing to $\tilde{r}'=\mathcal{M}\cdot s'$, the reco $\delta p_T$ forward-folded from wrong migration matrix model.
        $\mathcal{M}$ and $\mathcal{M'}$ are the migration matrices.
        Ratio $\tilde{r}'/r'$ is plotted in the bottom panel.
        (d): $s'$ comparing to $u'=\mathcal{M'}^{-1}\cdot r'$ and $\tilde{u}'=\mathcal{M}^{-1}\cdot r'$, the truth $\delta p_T$ unfolded from correct and wrong migration model. $\mathcal{M}^{-1}$ and $\mathcal{M}'^{-1}$ are migration matrices' pseudo-inverse calculated using singular value decomposition with a cutoff at 0.07. Bias $(\tilde{u}'-u')/s'$ is plotted in the bottom panel.
        For all of the above quantities, error bands and bars are MC statistical errors. In addition, the fractional neutrino interaction systematic error (GENIE systematics) and the MC statistical error of MINERvA's published $\delta p_T$ cross section \cite{Cai:2020PRD} are plotted in (b), (c)'s ratio subplot and (d)'s bias subplot as gray bands.
    }
    \label{fig:signal-reco-efficiency}
\end{figure*}

In this subsection, we test the BDT reweighting in a realistic use case.
In a real measurement using data, extracting the cross section measurement requires several steps, including estimations of neutrino flux, signal and background processes, detector efficiency and response, unfolding to correct for smearing, and so forth.
For this example, we use the BDT reweight to change the model for the signal process in a flux-weighted measurement of $\delta p_T$ from MINERvA's sample of QE-like events, and compare these to systematic and statistical uncertainties in the previously published analysis described in Ref~\cite{Cai:2020PRD}. Such a change would have little impact on the flux determination or backgrounds subtraction, but could change the efficiency because of differences in final-state kinematics, including the variation of the predicted cross section with neutrino energy, and because of changes in the migration between true and reconstructed quantities. Both of these effects are predicted from the signal model, and are convolved with the detector simulation.

The simulation for this analysis is taken from the FHC (neutrino beam) medium energy sample of $\nu_\mu$-carbon CCQE-like signal MC events from the MINERvA-Analysis-Toolkit (MAT) and MINERvA's open data product \cite{Messerly_2021, MINERvAopenData}. 
The neutrino interaction model in this data release is GENIE v2.12.6. We have dropped with unphysical elastic FSI events, as discussed earlier.
Signal events are required to have a final-state contains extactly one muon, $\ge1$ proton(s) with $T_p\ge50$MeV, any number of neutron(s), no photons, and no mesons.
The selection on reconstructed quantities follows MINERvA's CCQE-like reco selection, as discussed in Ref.~\cite{Cai:2020PRD}.
Events are divided into categories introduced in section $\ref{sec:event-categorization}$, and truth level quantities are fed to trained BDT reweighters that reweight these events to mimic GENIE v3's distributions.

Truth and reconstructed $\delta p_T$ distributions of 
the selected signal events, $s$ and $r$, are plotted in Figure \ref{fig:signal-reco-efficiency}(a) using the $\delta p_T$ binning from Ref.~\cite{Cai:2020PRD}. Primed quantities are these distributions with BDT reweight turned on. Ratios $s'/s$ and $r'/r$ show that the BDT reweight leads to large up-shifts in low $\delta p_T$ region and down-shifts in high $\delta p_T$ region. The shifts in reconstructed space are restrained compared to that in truth space.

Reconstruction selection efficiency is the number of selected events divided by total number of CCQE-like events in the bin.
If truth and reconstructed bins are denoted with subscripts $i$ and $j$, then $\delta p_T$ reco selection efficiency $\varphi$ is
\begin{equation}
    \varphi_i=\frac{s_i}{t_i},
\end{equation}
where $t$ is the $\delta p_T$ distribution of all CCQE-like events. Similarly, $\varphi_i'=s_i'/t_i'$. 
Efficiencies $\varphi',\varphi$, and ratio $\varphi'/\varphi$ are plotted in \ref{fig:signal-reco-efficiency}(b).  
The BDT reweight effects do change the efficiency as we might anticipate. The ratio plot in bottom panel shows that these shifts are mostly covered by the neutrino interaction fractional systematic error band of MINERvA's published $\delta p_T$ cross section's (GENIE systematics) \cite{Cai:2020PRD}, except the lower $\delta p_T$ bins and the last bin.

The migration matrices $\mathcal{M}$ and $\mathcal{M}'$ are 2D event-counting-histograms of truth versus reconstructed $\delta p_T$ bins where rows (indexed with $i$) are normalized. They satisfy
\begin{equation}
    r_j =\sum_i \mathcal{M}_{ij}s_i,
    r_j' =\sum_i \mathcal{M}'_{ij}s'_i.
\end{equation}
$\mathcal{M}_{ij}$ and $ \mathcal{M}'_{ij}$ are estimates of the probability that events from truth bin $i$ migrated to reconstructed bin $j$.
A bias in the result due to the change in the MC simulation is expected if the reconstructed $\delta p_T$ is forward-folded using the correct truth information but with the wrong migration matrix. To quantify this bias, we define
\begin{equation}
    \tilde{r}'_j =\sum_i \mathcal{M}_{ij}s_i'
\end{equation}
as the reconstructed $\delta p_T$ distributions extracted from BDT weighted truth $s'$ and unweighted migration matrix $\mathcal{M}$.
$\tilde{r}_j'$ and $r_j'$ are plotted in Figure \ref{fig:signal-reco-efficiency}(c), where the two curves are almost indistinguishable.
A closer look in the ratio $\tilde{r}'/r'$ suggests that shifts due to wrong migration model are small and almost well covered by the neutrino interaction systematic error band, signifying that the BDT reweight only has small effects in truth-to-reconstructed migration.

To inspect BDT reweight effects in unfolding, migration matrices' pseudo-inverse $\mathcal{M}^{-1}$ and $\mathcal{M'}^{-1}$ are derived using singular value decomposition (SVD) from $\mathcal{M}$ and $\mathcal{M'}$, with a singular value cutoff at 0.07.
Define
\begin{equation}
    u'_i =\sum_j \mathcal{M'}_{ij}^{-1}r_j',
    \tilde{u}'_i =\sum_j \mathcal{M}_{ij}^{-1}r_j'
\end{equation}
as the truth $\delta p_T$ distributions unfolded from $r'$, the weighted reconstructed $\delta p_T$, with correct and wrong migration matrix model. 
$u'$ and $\tilde{u}'$ are plotted in \ref{fig:signal-reco-efficiency}(d). A few fluctuations from truth $s'$ are seen, as unfolding from matrices' pseudo-inverse is unstable in nature.
Bias in the unfolded truth $(\tilde{u}'-u')/s'$ is plotted in the bottom panel, where we found that these fluctuations are likely statistical. They are mostly covered by the fractional statistical error band of MINERvA's published $\delta p_T$ cross section, except the first bin.  Here the statistical band is the appropriate measure because most of the differences not captured in the forward-folding study in \ref{fig:signal-reco-efficiency}(c) are statistical fluctuations in the unfolding.

As expected, most shifts in truth-to-reconstructed migration due to BDT reweight as seen by the forward-folding or unfolding from wrong model are covered by the gray systematic or statistical error bands, despite a few exceptions in efficiency ratio and unfolding bias.

We stress that the results of this study should not be interpreted as corrections to previous MINERvA measurements of these quantities. 
The reference models for those measurements were not default GENIE v2.12.6 as used in this paper, but rather tuned versions which were altered to better agree with distributions in the data.

\section{Conclusions}
In this study, we developed a generic method of multidimensional reweighting of generator predictions of GeV energy neutrino interaction samples that helps avoid heavy computation in MC generation and enable an efficient reuse of legacy data. The samples are reweighted using a boosted decision tree method to match quantities observable in a detector. We illustrated the method by using it to reweight the \minerva\ ME CCQE-like \numuC MC sample generated by GENIE v2.12.6 to match a sample generated by GENIE v3.04.00 AR23\_20i\_00\_000. The reweighting divides events into categories based on observable particle multiplicity to reduce the potential number of dimensions to consider. Reweighters used to estimate weights for a statistically independent GENIE v2 test sample demonstrate the ability reproduce the GENIE v3 AR23 sample's predictions, even for derived variables that were not directly part of the training. We also demonstrated that such a method can be used to reproduce measurements of efficiency as would be used in a cross section measurement in MINERvA.

For a complete application in MINERvA or other experiments, the method could be generalized to other categories of events, such as single pion production, and to predictions generated by other neutrino interaction generators, such as NuWro or NEUT.
Systematic uncertainties would need to be evaluated and implemented using the predictions of the target model, rather than the source model.  This is true even in the case that the source of an uncertainty is common in the two models, such as uncertainty of the axial form factor because correlations between events in the source and target models in variables not used in the tuning may not be preserved in reweighting.
Our procedure could evaluate systematic uncertainties in a target model by evaluating the changes in the target model for each uncertainty, and explicitly training the reweighting for each such variation. While the procedure would consume significantly more computing resources, we deem this approach as tractable since the training and storing of BDT reweight examples we showed in this paper is CPU-efficient and doesn't rely on extensive GPU clusters.

\begin{acknowledgments}
This document was prepared by members of the MINERvA Collaboration using the resources of the Fermi National Accelerator Laboratory (Fermilab), a U.S. Department of Energy, Office of Science, HEP User Facility. Fermilab is managed by Fermi Research Alliance, LLC (FRA), acting under Contract No. DE-AC02-07CH11359.
Support for participating scientists was provided by NSF and DOE (USA); by CAPES and CNPq (Brazil); by CoNaCyT (Mexico); by ANID PIA / APOYO AFB180002, CONICYT PIA ACT1413, and Fondecyt 3170845 and 11130133 (Chile); by CONCYTEC (Consejo Nacional de Ciencia, Tecnolog\'ia e Innovaci\'on Tecnol\'ogica), DGI-PUCP (Direcci\'on de Gesti\'on de la Investigaci\'on  - Pontificia Universidad Cat\'olica del Peru), and VRI-UNI (Vice-Rectorate for Research of National University of Engineering) (Peru); NCN Opus Grant No. 2016/21/B/ST2/01092 (Poland); by Science and Technology Facilities Council (UK); by EU Horizon 2020 Marie Skłodowska-Curie Action; by a Cottrell Postdoctoral Fellowship from the Research Corporation for Scientific Advancement; by an Imperial College London President's PhD Scholarship.

\end{acknowledgments}

\FloatBarrier

\appendix

\section{Reweight Results of Individual Categories}\label{appx:figures}
\begin{figure*}[tp]
    \centering
    \includegraphics[width=1\linewidth]{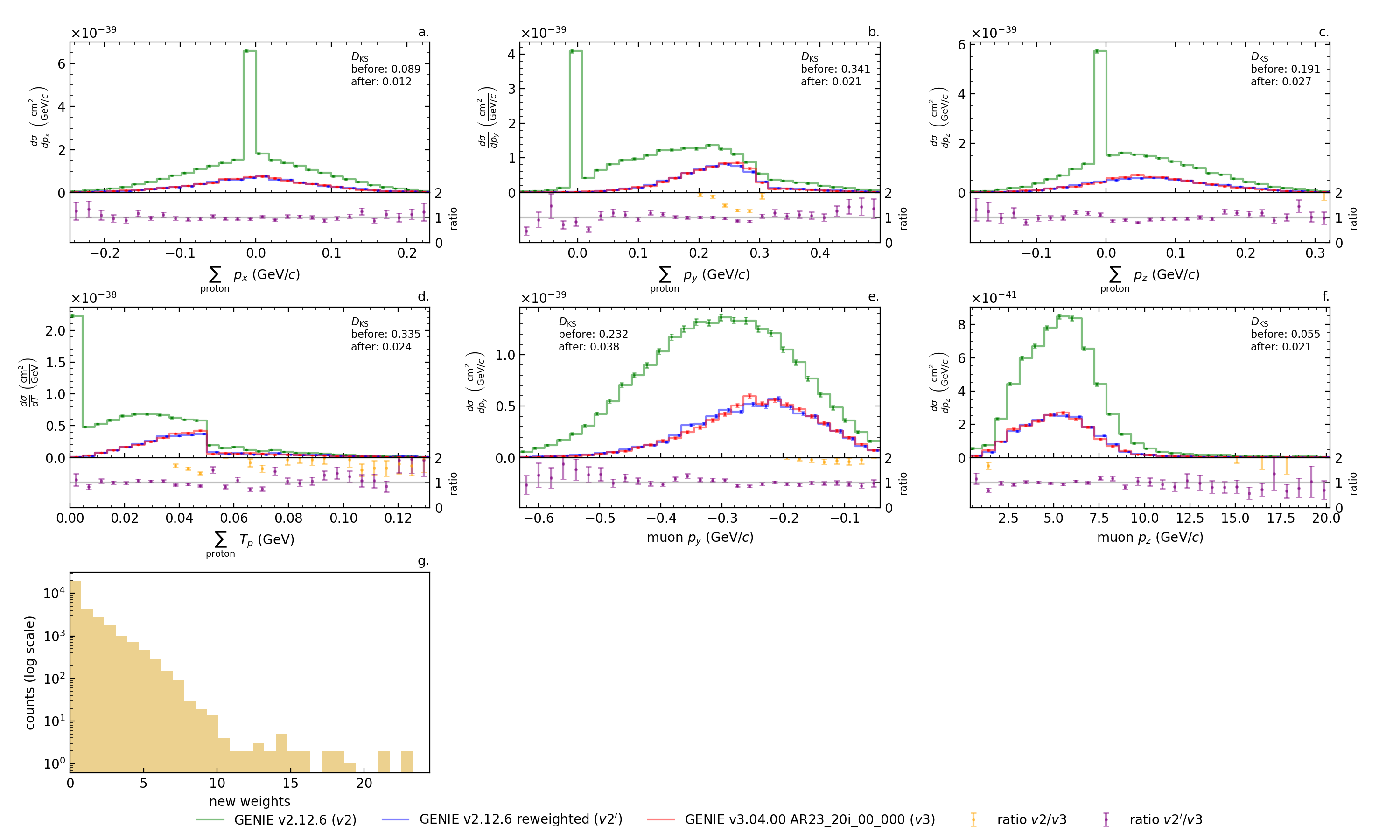}
    \caption{
        Differential cross sections of category 0p0n are plotted with respect to calorimetric momenta $\sum p_x, \sum p_y, \sum p_z,$ and energy $\sum T_p$ summed over all final-state protons (a, b, c, d); and $\mu^-$ $p_y,p_z$ (e, f).
        Histogram of weights (g) is plotted in log scale.
        Green: test sample \genieTwo~(v2). Blue: reweighted test sample (v2$'$). Red: target sample \genieThree~(v3).
        cross section ratios of v2 and v2$'$ comparing to v3 are plotted under each histogram, in yellow and purple respectively.
        K-S test statistic $D_{\text{KS}}$ before (v2 comparing to v3) and after (v2$'$ comparing to v3) reweighting is printed on each histogram.
    }
    \label{fig:0p0n-test}
\end{figure*}

\begin{figure*}[tp]
    \centering
    \includegraphics[width=1\linewidth]{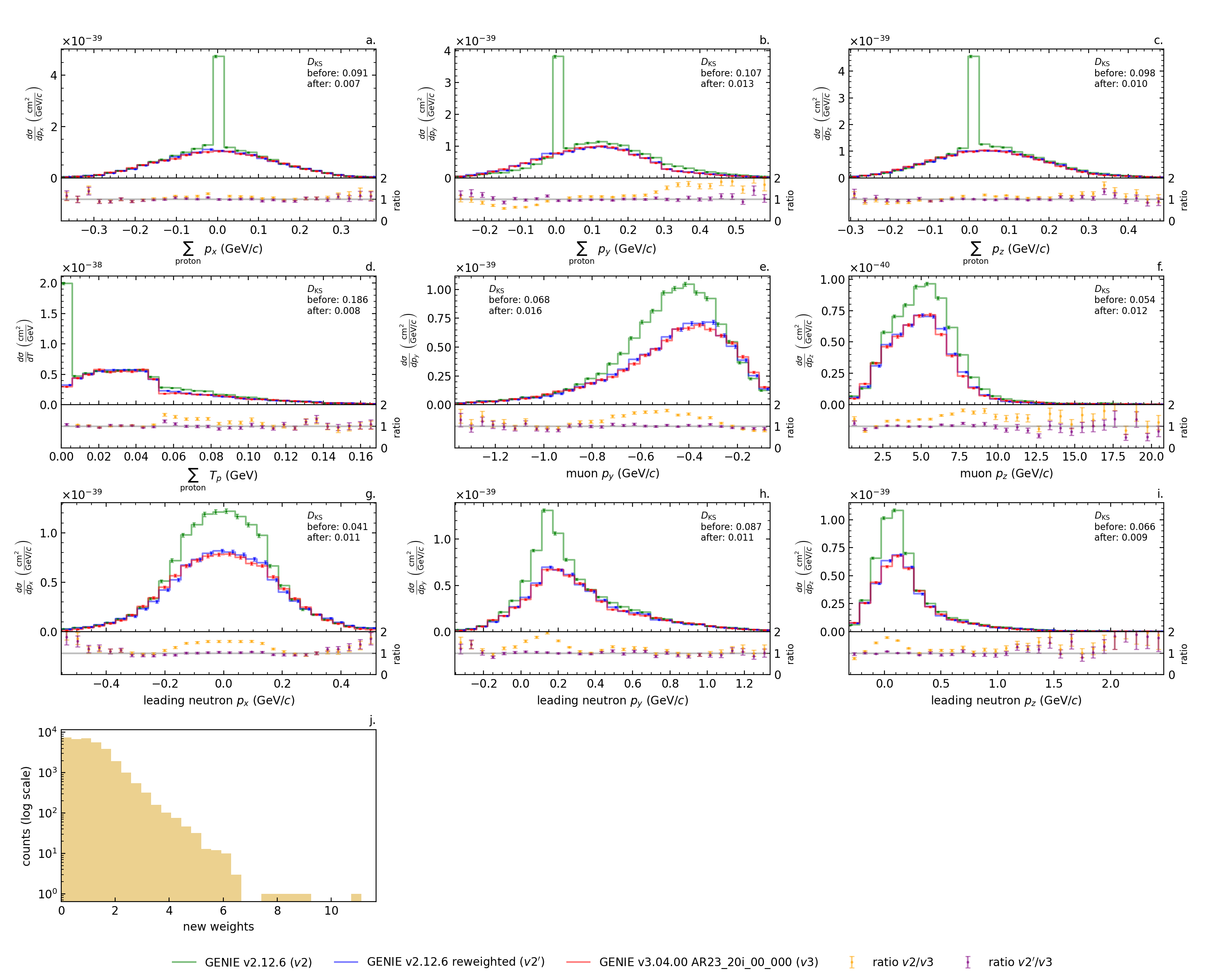}
    \caption{
        Differential cross sections of category 0pNn are plotted with respect to calorimetric momenta $\sum p_x, \sum p_y, \sum p_z,$ and energy $\sum T_p$ summed over all final-state protons (a, b, c, d); $\mu^-$ $p_y,p_z$ (e, f); and leading neutron $p_x, p_y, p_z$ (g, h, i) .
        Histogram of weights (j) is plotted in log scale.
        Green: test sample \genieTwo~(v2). Blue: reweighted test sample (v2$'$). Red: target sample \genieThree~(v3).
        cross section ratios of v2 and v2$'$ comparing to v3 are plotted under each histogram, in yellow and purple respectively.
        K-S test statistic $D_{\text{KS}}$ before (v2 comparing to v3) and after (v2$'$ comparing to v3) reweighting is printed on each histogram.
    }
    \label{fig:0pNn-test}
\end{figure*}

\begin{figure*}[tp]
    \centering
    \includegraphics[width=1\linewidth]{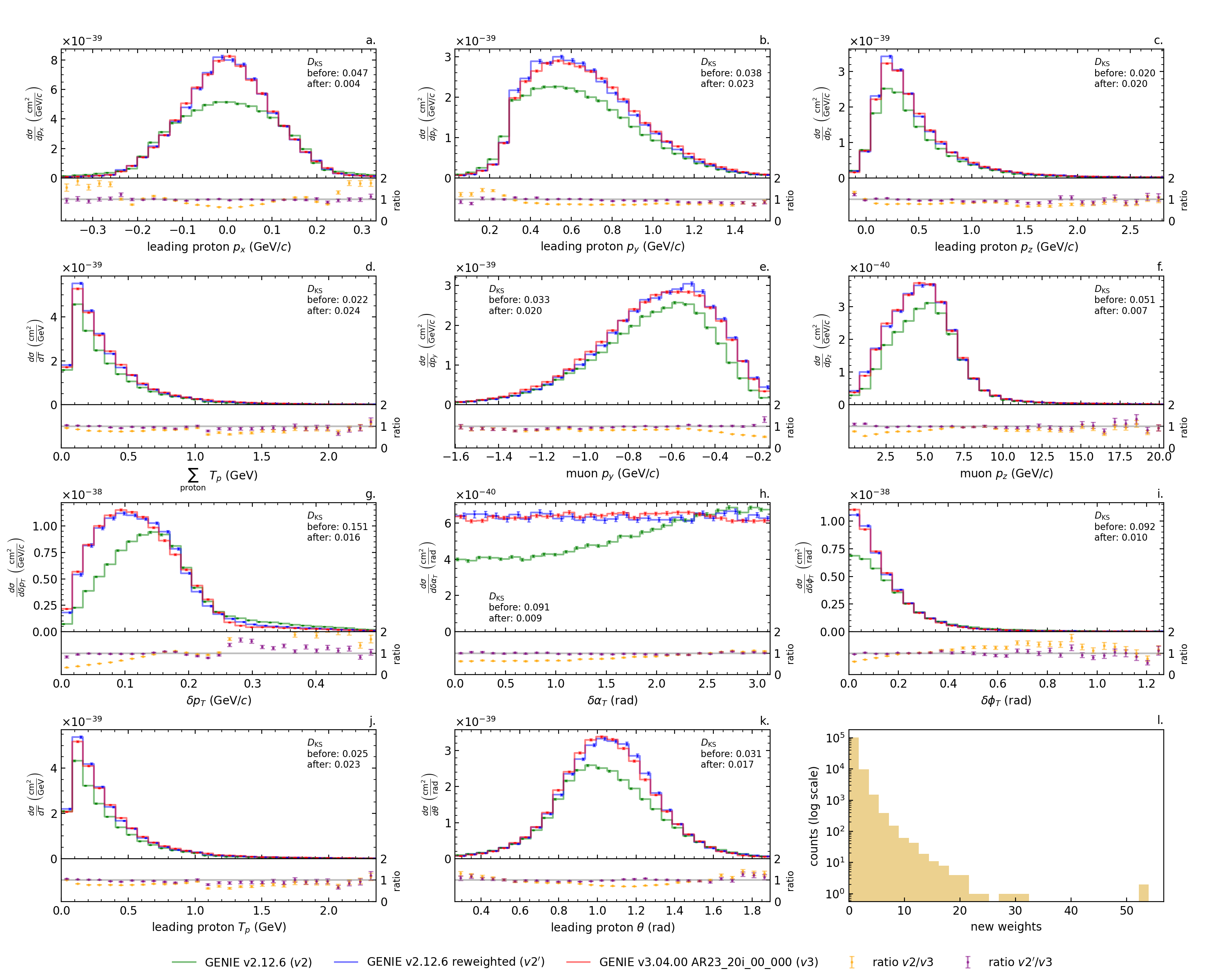}
    \caption{
        Differential cross sections of category 1p0n are plotted with respect to leading proton $p_x, p_y,p_z$ (a, b, c); calorimetric energy $\sum T_p$ (d); $\mu^-\ p_y,p_z$ (e, f); TKI variables $\delta p_T,\delta \alpha_T,\delta \phi_T$ (g, h, i); and leading proton $T_p,\theta$ (j, k).
        Histogram of weights (l) is plotted in log scale.
        Green: test sample \genieTwo~(v2). Blue: reweighted test sample (v2$'$). Red: target sample \genieThree~(v3).
        cross section ratios of v2 and v2$'$ comparing to v3 are plotted under each histogram, in yellow and purple respectively.
        K-S test statistic $D_{\text{KS}}$ before (v2 comparing to v3) and after (v2$'$ comparing to v3) reweighting is printed on each histogram.
    }
    \label{fig:1p0n-test}
\end{figure*}

\begin{figure*}[tp]
    \centering
    \includegraphics[width=1\linewidth]{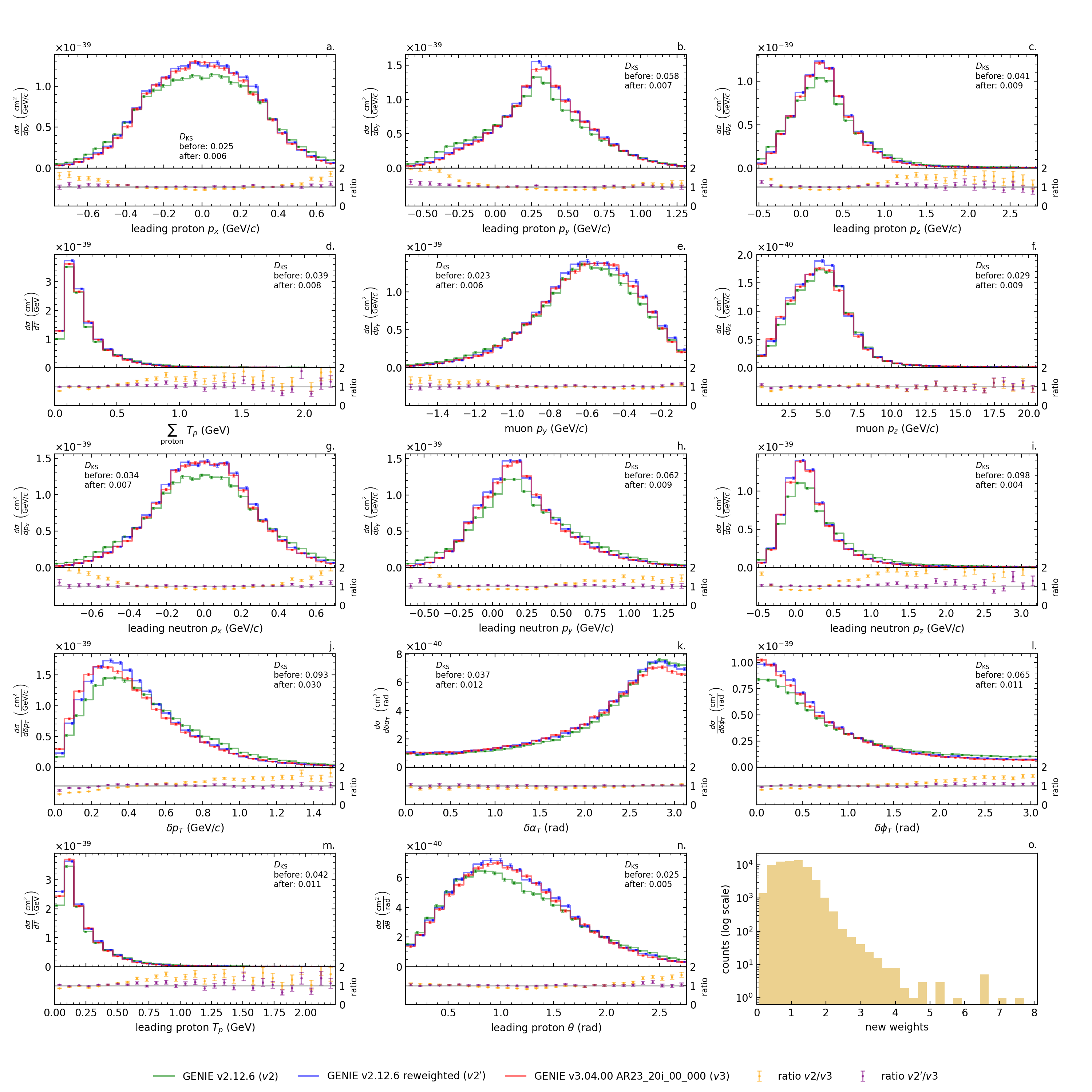}
    \caption{
       Differential cross sections of category 1pNn are plotted with respect to leading proton $p_x, p_y,p_z$ (a, b, c); calorimetric energy $\sum T_p$ (d); $\mu^-\ p_y,p_z$ (e, f); leading neutron $p_x,p_y,p_z$ (g, h, i); TKI variables $\delta p_T,\delta \alpha_T,\delta \phi_T$ (j, k, l); and leading proton $T_p,\theta$ (m, n).
        Histogram of weights (o) is plotted in log scale.
        Green: test sample \genieTwo~(v2). Blue: reweighted test sample (v2$'$). Red: target sample \genieThree~(v3).
        cross section ratios of v2 and v2$'$ comparing to v3 are plotted under each histogram, in yellow and purple respectively.
        K-S test statistic $D_{\text{KS}}$ before (v2 comparing to v3) and after (v2$'$ comparing to v3) reweighting is printed on each histogram.
        }
    \label{fig:1pNn-test}
\end{figure*}

\begin{figure*}[tp]
    \centering
    \includegraphics[width=1\linewidth]{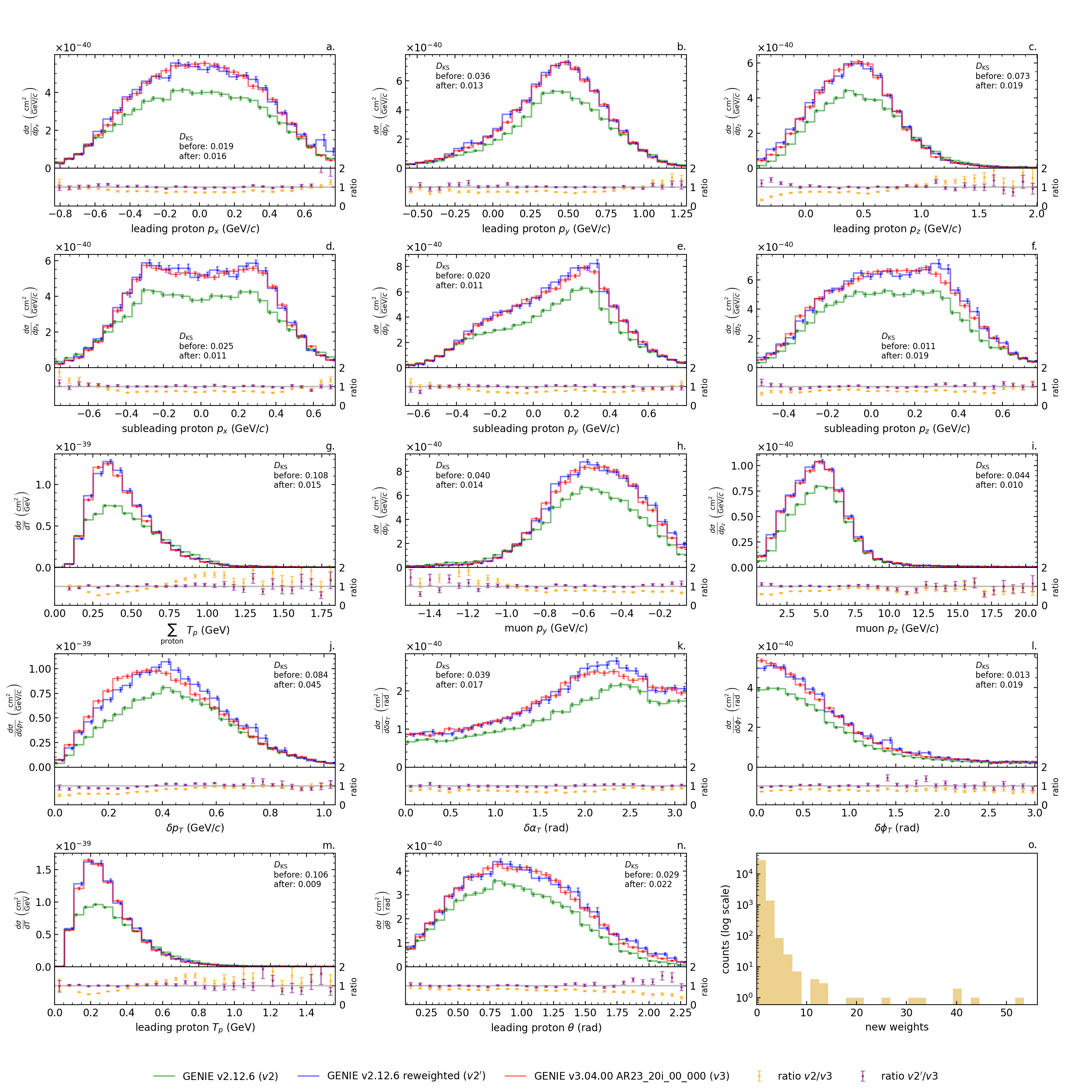}
    \caption{
        Differential cross sections of category 2p0n are plotted with respect to leading proton $p_x, p_y,p_z$ (a, b, c); second leading proton $p_x,p_y,p_z$ (d, e, f); calorimetric energy $\sum T_p$ (g); $\mu^-\ p_y,p_z$ (h, i); TKI variables $\delta p_T,\delta \alpha_T,\delta \phi_T$ (j, k, l); and leading proton $T_p,\theta$ (m, n).
        Histogram of weights (o) is plotted in log scale.
        Green: test sample \genieTwo~(v2). Blue: reweighted test sample (v2$'$). Red: target sample \genieThree~(v3).
        cross section ratios of v2 and v2$'$ comparing to v3 are plotted under each histogram, in yellow and purple respectively.
        K-S test statistic $D_{\text{KS}}$ before (v2 comparing to v3) and after (v2$'$ comparing to v3) reweighting is printed on each histogram.
        }
    \label{fig:2p0n-test}
\end{figure*}

\begin{figure*}[tp]
    \centering
    \includegraphics[height=0.85\textheight,width=1\linewidth]{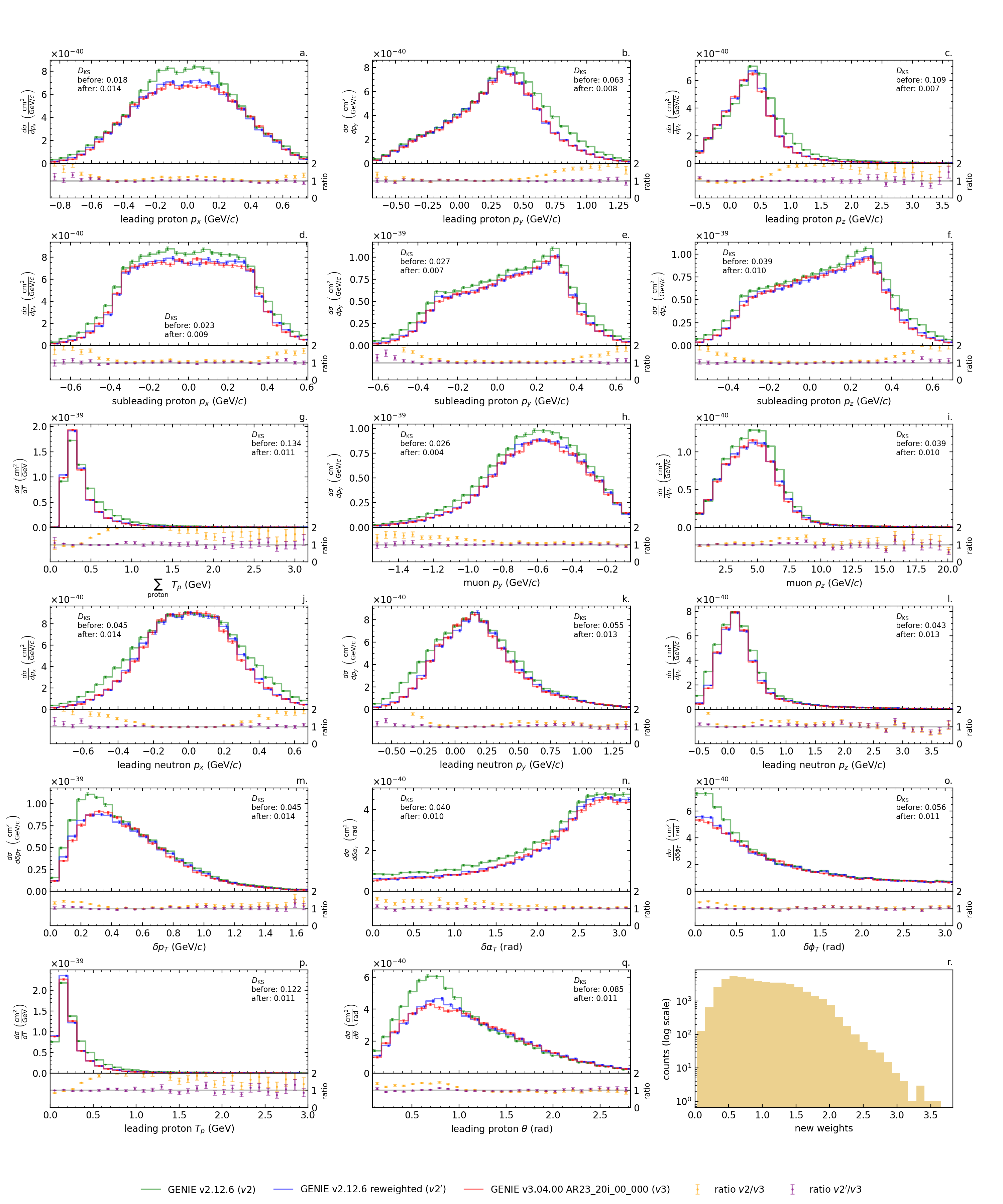}
    \caption{
        Differential cross sections of category 2pNn are plotted with respect to leading proton $p_x, p_y,p_z$ (a, b, c); 
        second leading proton $p_x,p_y,p_z$ (d, e, f); 
        calorimetric energy $\sum T_p$ (g); 
        $\mu^-\ p_y,p_z$ (h, i);
        leading neutron $p_x, p_y,p_z$ (j, k, l);
        TKI variables $\delta p_T,\delta \alpha_T,\delta \phi_T$ (m, n, o); 
        and leading proton $T_p,\theta$ (p, q).
        Histogram of weights (r) is plotted in log scale.
        Green: test sample \genieTwo~(v2). Blue: reweighted test sample (v2$'$). Red: target sample \genieThree~(v3).
        cross section ratios of v2 and v2$'$ comparing to v3 are plotted under each histogram, in yellow and purple respectively.
        K-S test statistic $D_{\text{KS}}$ before (v2 comparing to v3) and after (v2$'$ comparing to v3) reweighting is printed on each histogram.
        }
    \label{fig:2pNn-test}
\end{figure*}

\begin{figure*}[tp]
    \centering
    \includegraphics[width=1\linewidth]{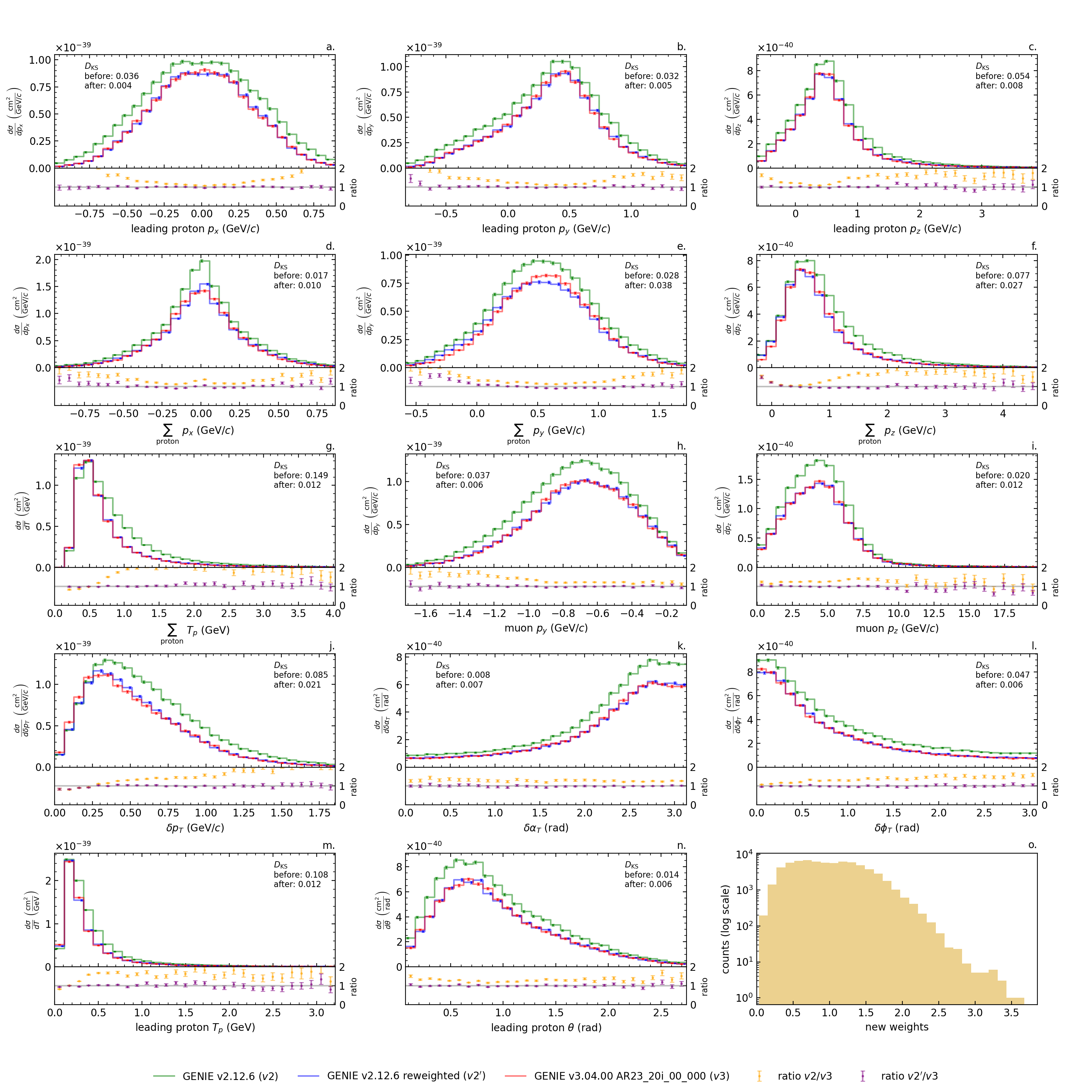}
    \caption{
        Differential cross sections of category ``others'' are plotted with respect to leading proton $p_x, p_y,p_z$ (a, b, c); 
        calorimetric momenta $\sum p_x, \sum p_y, \sum p_z,$ and energy $\sum T_p$ summed over all final-state protons (d, e, f, g);
        $\mu^-\ p_y,p_z$ (h, i);
        TKI variables $\delta p_T,\delta \alpha_T,\delta \phi_T$ (j, k, l); 
        and leading proton $T_p,\theta$ (m, n).
        Histogram of weights (o) is plotted in log scale.
        Green: test sample \genieTwo~(v2). Blue: reweighted test sample (v2$'$). Red: target sample \genieThree~(v3).
        cross section ratios of v2 and v2$'$ comparing to v3 are plotted under each histogram, in yellow and purple respectively.
        K-S test statistic $D_{\text{KS}}$ before (v2 comparing to v3) and after (v2$'$ comparing to v3) reweighting is printed on each histogram.
    }
    \label{fig:others-test}
\end{figure*}

This appendix shows the reweight results  for events from the seven individual categories listed in Table \ref{tab:categories-variables}. Differential cross sections with respect to selected final-state kinematic variables of test sample \genieTwo, reweighted test sample, and target sample \genieThree\ are plotted in Figure \ref{fig:0p0n-test} through \ref{fig:others-test}.

\begin{figure*}[tp]
    \centering
    \includegraphics[width=1\linewidth]{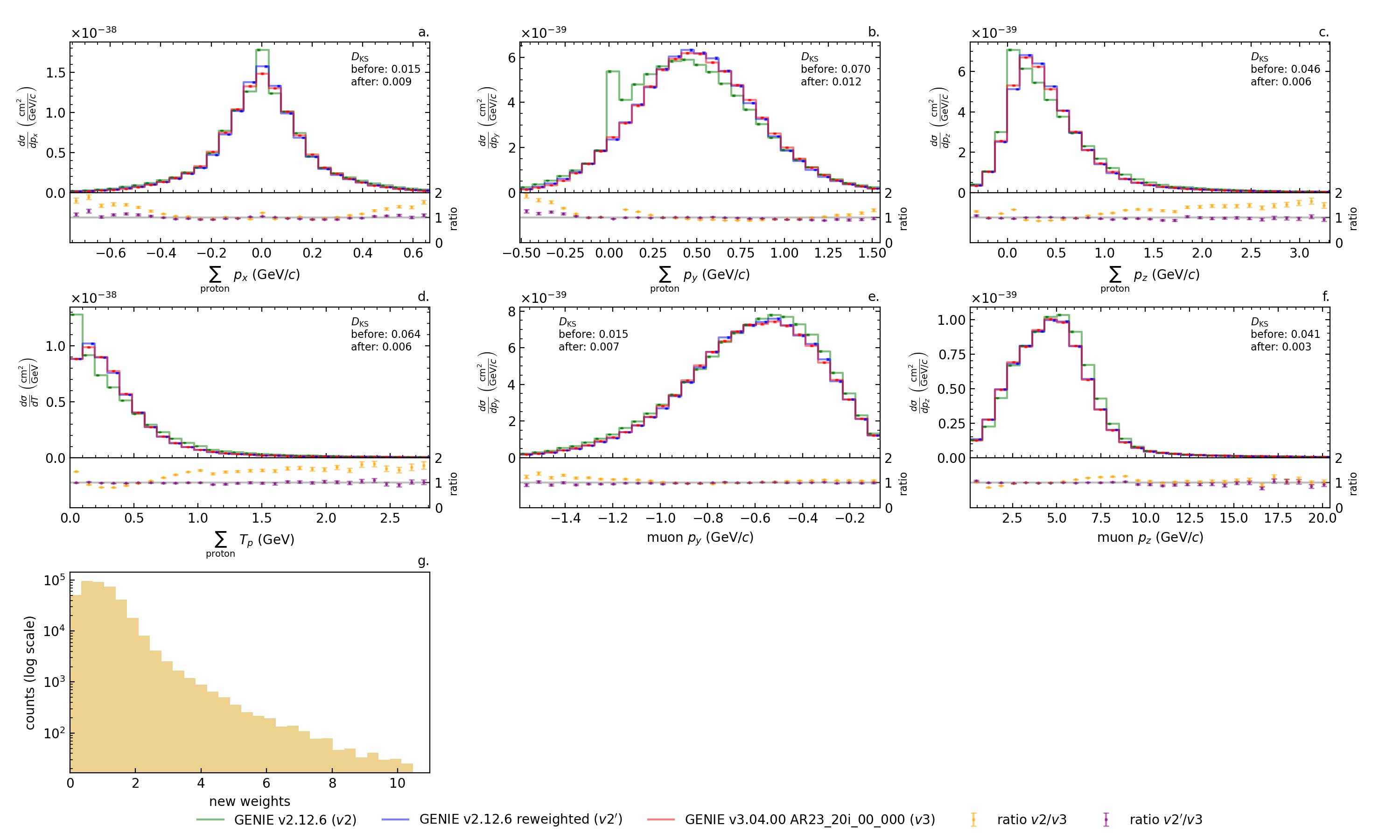}
    \caption{
        Companion plots of Figure \ref{fig:combined-sumProton-test} where GENIE v2.12.6 events with weight $>10$ are dropped.
    }
    \label{fig:combined-sumProton-test-weight-0-10}
\end{figure*}

\begin{figure*}[tp]
    \centering
    \includegraphics[width=1\linewidth]{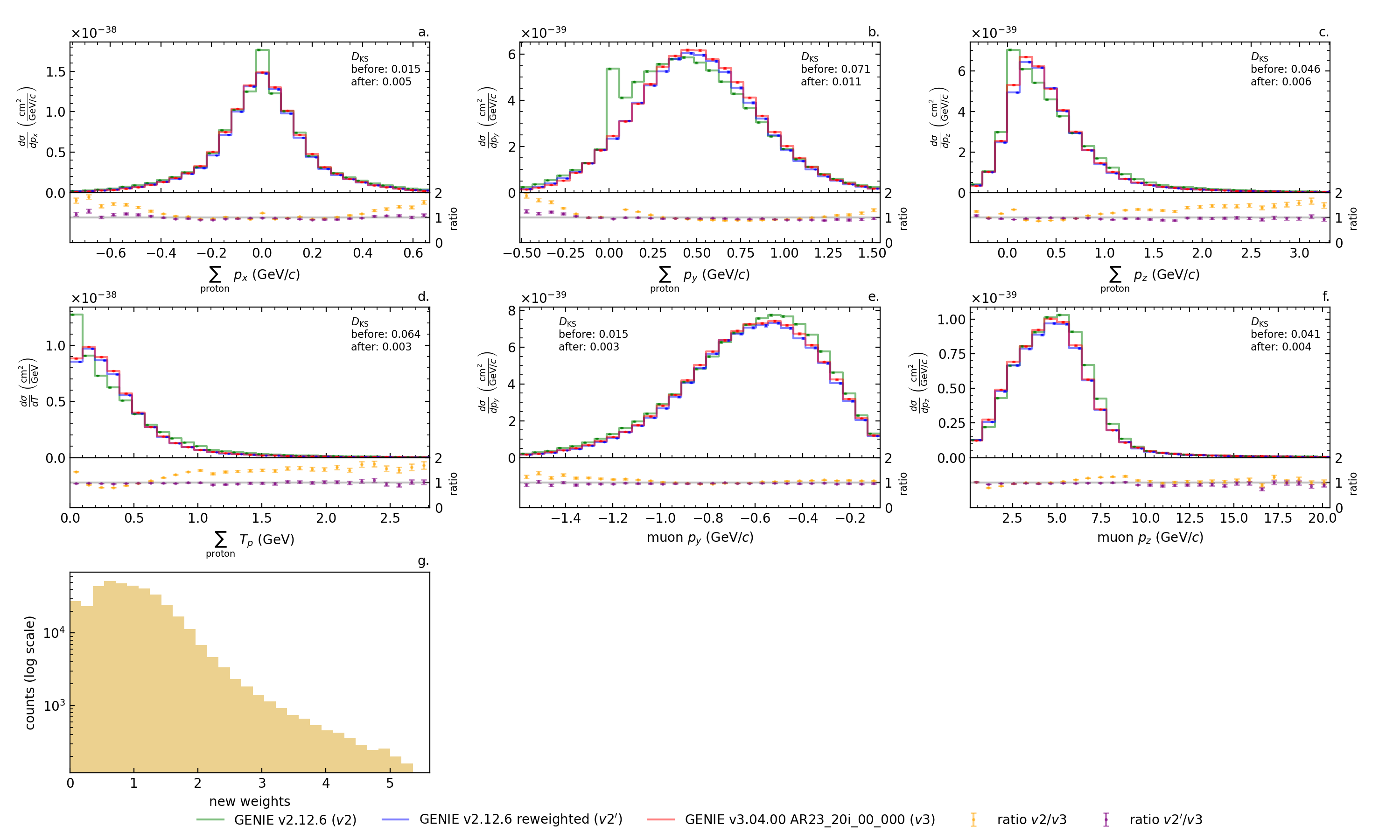}
    \caption{
        Companion plots of Figure \ref{fig:combined-sumProton-test} where GENIE v2.12.6 events with weight $>5$ are dropped.
    }
    \label{fig:combined-sumProton-test-weight-0-5}
\end{figure*}

\begin{figure*}[tp]
    \centering
    \includegraphics[width=1\linewidth]{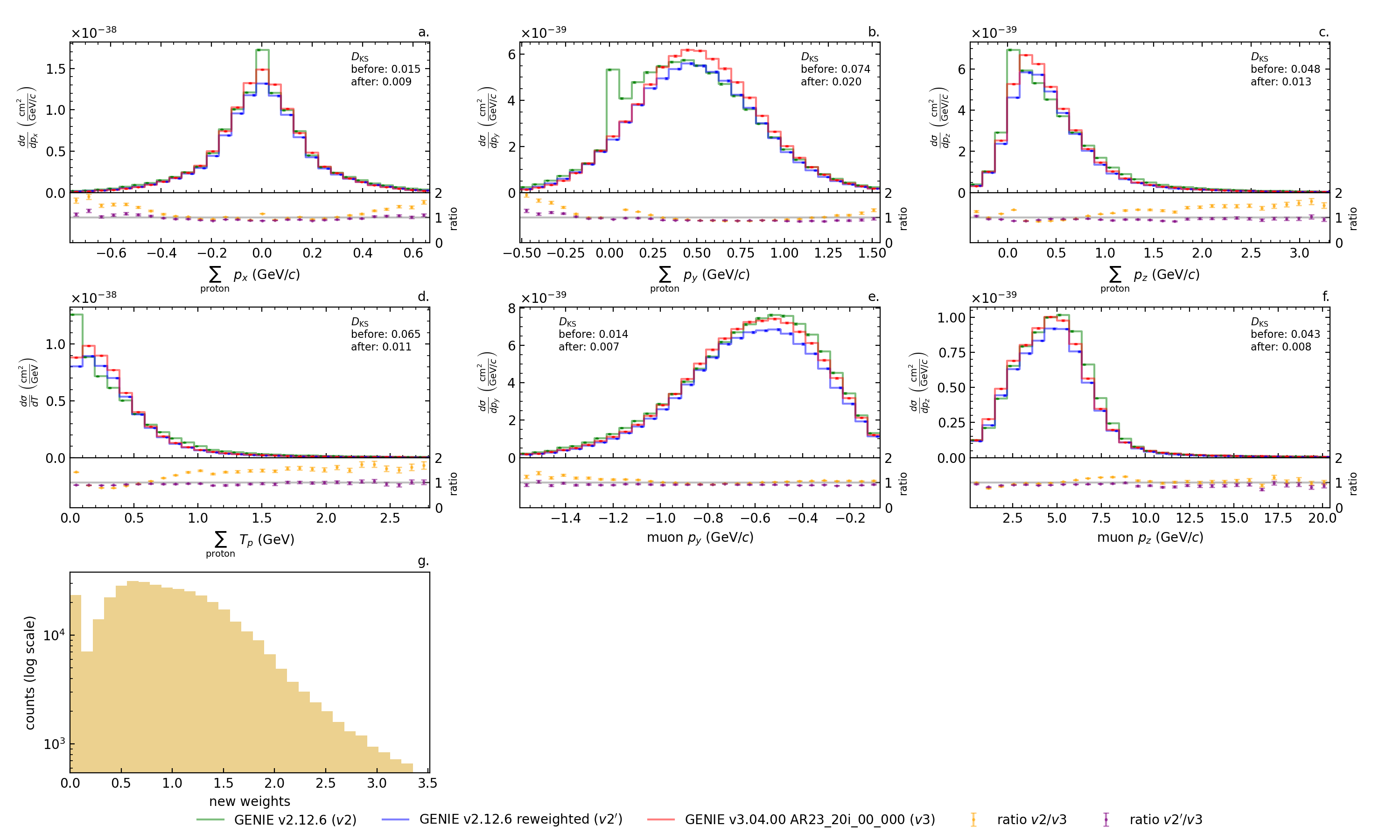}
    \caption{
        Companion plots of Figure \ref{fig:combined-sumProton-test} where GENIE v2.12.6 events with weight $>3$ are dropped.
    }
    \label{fig:combined-sumProton-test-weight-0-3}
\end{figure*}

For all figures in this study, GENIE v2.12.6 events with predicted weights $>100$ are thrown out.
We investigated in these events briefly and concluded that they occur in rare, and thus low statistics, kinematic regions, and it is therefore safe to discard them. For example, some of them are DIS events in the GENIE classification without any out going pion and so fall into our CCQE selection even though such events are unusual.
To further demonstrate our point, we investigated the impact of constraining range of weights.
Companion plots to Figure \ref{fig:combined-sumProton-test} are produced to show the impact of discarding large weight events, where cross-secitons of events with weight in range $[0,10]$, $[0,5]$, and $[0,3]$ are plotted in Figure \ref{fig:combined-sumProton-test-weight-0-10}, \ref{fig:combined-sumProton-test-weight-0-5}, and \ref{fig:combined-sumProton-test-weight-0-3}, respectively. No visible differences are seen in either cross sections or ratio plots comparing to Fig.~\ref{fig:combined-sumProton-test}, although small changes in the test statistic $D_{KS}$ are observed.

\section{GENIE Processes that Lead to Zero Available Energy in Carbon}\label{appx:spikes-GENIE2}
Most of the production of zero proton kinetic energy events is from a specific combination of choices in GENIE v2. 
GENIE v2 subtracts 25 MeV from each nucleon when there is one or two nucleons in the final-state, after FSI if any was chosen.
This accounts for the energy cost to remove these nucleons from the nucleus.
The most common example for carbon is when the proton undergoing a single nucleon knockout reaction, with the resulting energy shared between two nucleons (say a pn pair) in the end.
When that proton has less than 25 MeV after the sharing, the subtraction means the proton is produced exactly at rest. This happens for about 4\% of QE events at MINERvA energies in GENIE v2, making a spike in the distribution.
This 25 MeV subtraction is not applied in GENIE v3.
There is another FSI process that divides the proton energy among three or more nucleons, with a small probability to produce only neutrons, also leading to exactly zero proton momentum and kinetic energy. 
This process is the same for GENIE v2 and GENIE v3, there is no 25 MeV subtraction in either case.
Because this second case happens for only 0.1\% of QE events, no spike is visible unless the distribution is extremely finely binned.

The ``shelf'' in the proton KE plot is related to this 25 MeV removal energy subtraction when there is no FSI process simulated. 
Such protons (in neutrino QE mode) start with at least 25 MeV based on the minimum energy transfer coded in and the required Q-value of the reaction.
Then in the GENIE v2 carbon case a quanta of 25 MeV is subtracted as described above, leading to near zero kinetic energy, even with no FSI energy sharing with other nucleons.
The situation is similar in GENIE v3 except for the subtraction, leading to a no-FSI spectrum with that 25 MeV of kinetic energy, while the FSI process continues to produce proton KE events down to rest.
For CC interactions in non-isoscalar nuclei such as Pb, the difference between proton and neutron removal energy can induce a spike at zero KE even for GENIE v3.

In the weighting procedure, the BDT simply weights the population in the spike to zero, and weights down events in the shelf, in order to describe the GENIE v3 population.
However, the are events in data down practically to zero energy, so there is limited utility in using GENIE v3 (via reweighting or directly) to describe the QE hadronic energy distribution.
In some future version of GENIE, this deficiency will be solved.
For MINERvA’s fully simulated GENIE v2 events, one could imagine an ad-hoc adjustment to the events in the shelf to now weight them down as strongly even when producing the other features of the GENIE v3 model predictions.
Such a strategy would require similarly careful interpretation of the prediction as using
GENIE v3 directly.

\bibliography{main}

\end{document}